\documentclass[11pt]{article}

\usepackage{amssymb}
\usepackage{graphics,graphicx}
\usepackage{t1enc}
\usepackage[english]{babel}

\def\theequation{\arabic{section}.\arabic{equation}}
\title{Light-cone}

\begin{document}

\begin{titlepage}

\begin{flushright}
                Preprint TPJU-2/2002\\
                February 2002  \\
                hep-th/0202051
\end{flushright}
\bigskip

\begin{center}
{\LARGE \bf Joining--splitting interaction of non-critical
string\\}
\end{center}
\bigskip

\begin{center}
    {\bf
    Leszek Hadasz}\footnote{e-mail: hadasz@th.if.uj.edu.pl} \\
    M. Smoluchowski Institute of Physics \\
    Jagiellonian University, \\
    Reymonta 4, 30-059 Krak\'ow, Poland \\
\vskip 5mm
    {\bf
    Zbigniew Jask\'{o}lski}\footnote{e-mail: Z.Jaskolski@if.uz.zgora.pl }\\
    Physics Institute\\
    University of Zielona G\'{o}ra\\
    pl. S{\l}owia\'{n}ski 6,
    65-069 Zielona G\'{o}ra, Poland

\end{center}

\vskip 1cm

\begin{abstract}
The joining--splitting interaction of non-critical bosonic string
is analyzed in the light-cone formulation. The Mandelstam method
of constructing tree string amplitudes is extended to the bosonic
massive string models of the discrete series. The general
properties of the Liouville longitudinal excitations which are
necessary and sufficient for the Lorentz covariance of
the light-cone amplitudes are derived.
The results suggest that the covariant and the light-cone approach are equivalent also
in the non-critical dimensions. Some aspects of unitarity of interacting non-critical massive string
theory are discussed.
\end{abstract}
\bigskip\bigskip

\thispagestyle{empty}
\end{titlepage}

\section{Introduction}

It has been well known since early days of string theory that the
covariant quantization of the free Nambu-Goto string
\cite{brower72} and the Ramond-Neveu-Schwarz fermionic string
\cite{schwarz72,brower73} leads in non-critical dimensions to
consistent quantum models with longitudinal excitations. The
relevance of the (super)-Liouville theory for a proper description
of these extra degrees of freedom was first pointed out by
Polyakov in his celebrated papers on conformal anomaly in string
theories \cite{poly81,poly81b}. The free string models obtained by
adding the (super)-Liouville sectors were first analyzed by
Marnelius \cite{marnelius}. More recently  the no-ghost theorems
for these models were derived \cite{haja,hajaos} yielding one
continuous and one discrete series of free non-critical string
models. In particular it was shown that the non-critical
Nambu-Goto string and the non-critical RNS string are members of
the corresponding discrete series. In all cases the first excited
state is massive which justifies the name {\it massive string} for
all these models. It was subsequently shown that massive strings
admit the light-cone formulation which can be used to analyze
their spin content \cite{dahaja98,dahaja99,daja00}. In contrast to
the critical strings where the number of the tachyon-free models
is strongly limited, the GSO projection  yields  a large class of
tachyon-free massive strings
 with rich mass and spin spectra \cite{daja00}.
None of these models contains massless states with spin
greater than 1. This makes them   good candidates for matter fields
in an effective description of low energy QCD.

In the present paper we shall discuss  some aspects of introducing
interactions for bosonic massive strings of the discrete series.
The solution of the physical state conditions in terms of DDF
operators yields in this case three types  of physical degrees of
freedom \cite{haja}: the transverse excitations described by the
$d-2$ tensor power of the 2-dim scalar field Fock space, the {\it
Brower longitudinal} excitations described by a unitary Verma
module with the central charge $0\leqslant c^{\rm
\scriptscriptstyle B}<1$, and the {\it Liouville longitudinal}
excitations described by the Verma module with the central charge
$1< c^{\rm \scriptscriptstyle L}=1+48\beta <25$ and the highest
weight $(2\beta,2\beta)$. Each of these models describes a free
string propagating in the flat $d$-dimensional Minkowski target
space, only the structure of the longitudinal sector changes along
the series. A special case where the Brower longitudinal
excitations drop out ($c^{\rm \scriptscriptstyle B}=0$) and the
longitudinal sector consists solely of the Liouville excitations
corresponds to the Nambu-Goto non-critical string\footnote{ In
this case the Liouville longitudinal sector of the massive string
corresponds to the Brower longitudinal sector of the non-critical
Nambu-Goto string. The name {\it Brower longitudinal excitations}
is motivated by the way in which they
 arise in the covariant quantization of  the massive string  --- it is
essentially the same as in the case of Brower longitudinal excitations in the
 non-critical Nabu-Goto string.
}. Our choice  to consider the
whole series is motivated by the properties of the Brower excitations which provide a convenient
``intermediate step'' between the properties of the transverse, and of the
Liouville longitudinal sector.

 In the case of the critical string there are two
strategies to construct interacting theory. The first one is the
Polyakov covariant approach which can be seen as a modern
covariant version of the old dual model construction \cite{dual}.
In this approach the on-shell string amplitudes  are defined in
terms of correlators of the 2-dim conformal field theory
integrated over moduli spaces of corresponding Riemann surfaces
\cite{dhph}. The second approach, introduced by Mandelstam
\cite{Mandelstam73,Mandelstam74,Mandelstam86}, is based on the
light-cone formulation of the free string.
In this approach the unitarity of the S-matrix
is manifest while the Lorentz covariance requires a (non-trivial)
proof \cite{Mandelstam73,Mandelstam74,Sin88,kisa}. The light-cone formulation is well developed only for
critical strings in the flat Minkowski background. In this case it is equivalent to the covariant
approach \cite{gw}  what provides a proof of the unitarity of the
Polyakov formulation.

Once a free string theory is known the problem of calculating string amplitudes
reduces in both formulations to the problem of extending the 2-dim theory describing the string degrees of
freedom from the cylinder to an arbitrary Riemann surface.
In the simplest case of tree amplitudes this is an extension of the
theory from the sphere with 2 punctures to the sphere with arbitrary
number of punctures. In the standard CFT such extension is straightforward.
Indeed the puncture with an associated external state can be replaced
by an insertion of a local operator what results in a correlator on a sphere
with no punctures. Thus there is no real difference between the free theory on the cylinder and the theory
on an arbitrary
punctured sphere.
This is not necessarily the case in the Liouville sector
where the operator-state correspondence is subtle and still not fully
understood \cite{sei,gm,Teschner:2001rv}.

Within the covariant approach one can define the massive string amplitudes by a straightforward
application of the critical string construction.
The transverse sector is included in the tensor product of $d$ copies of the free scalar CFT.
The Brower sector can be extended to a CFT on an arbitrary Riemann surface by combining
several unitary Verma modules into the Hilbert space of an appropriate minimal model.
Following this line of reasoning one should then extend the Liouville sector to the CFT
with the central charge $1<c^{\rm \scriptscriptstyle L}=1+48\beta < 25$  and with a single
$(2\beta,2\beta)$-conformal family.

The structure of this theory differs in two respects from the
structure of standard CFT of other sectors. First, the $(0,0)$-family does not belong to
its spectrum ({\em i.e.} there is no $PSL(2,\mathbb{C})$-invariant vacuum).
Second, as can be easily inferred from the Vafa condition
\cite{Vafa88,Lewellen89}, with only one conformal family
the standard OPE for the punctures does not hold \cite{hadjas}.
This means in particular that the correlators do not factorize
on the free spectrum (in  commonly adopted
approaches to the quantum Liouville theory the spectrum required by factorization is actually
continuous \cite{sei}).

The main problem in calculating tree amplitudes of string models under consideration is thus
a construction of a consistent Liouville theory on arbitrary punctured sphere.
Such  theory is well understood in the so
called weak coupling regime, {\em i.e.} for the central charge of the Liouville sector
in the range $c^{\rm \scriptscriptstyle L}\leqslant 1$ or $25
\leqslant c^{\rm \scriptscriptstyle L}$ \cite{ddk,sei,gm}.
Indeed in this case the perturbative continuous formulation is
in a perfect agreement with the topological gravity and the matrix model results \cite{gm}.
However, in the strong coupling region  $1< c^{\rm \scriptscriptstyle L}<25$
it leads  to complex critical exponents \cite{sei,gm}.
This phenomenon  can be interpreted as a manifestation
of the tachyon instability \cite{sei}.

One way to evade this so called $c=1$ barrier was proposed by Gervais  \cite{Gervais:1990be}.
In this approach
the continuous Liouville spectrum is truncated to a discrete subset which enters both the
external states and the factorization.
  However, due to technicalities involved
\cite{Gervais:1994ec, Gervais:1996vp}, it is still not clear
whether this approach may lead to consistent string models in
non-critical dimensions.
Another possibility is related to the conjecture of Al. and A. Zomolodchikov
\cite{Zamolodchikov:1995aa} that the continuum spectrum along with
the 3-point functions proposed by Otto and Dorn \cite{Dorn:1994xn}
satisfy the bootstrap consistency conditions. This conjecture was
recently proved by Ponsot and Teschner
\cite{Ponsot:1999uf,Ponsot:2000mt} in the weak coupling regime
$c>25$. By the analytic continuation argument
it is believed to hold in the strong coupling regime
$1<c<25$ as well
\cite{Teschner:2001rv}. Finally there exists
 the  geometric approach to the 2-dim
quantum gravity originally proposed by Polyakov \cite{Polyakov82}
and further developed  by Takhtajan
\cite{Takhtajan:1993vt,Takhtajan:zi,Takhtajan:1994vt,Takhtajan:1995fd}.
According to this approach the Liouville correlation functions
are defined in terms of path integral over conformal class of Riemannian metrics
in which  the vertex operator insertions are replaced by
 metric singularities at the insertion points.
None of these approaches is developed well enough to provide correlation functions
even in the simplest case of punctured spheres.

Even though any ultimate solution to the Liouville theory has not been found yet it is still reasonable to
analyze the relation between its properties and the properties of the resulting non-critical string.
In the present paper we address two problem of this kind: the  construction of
the light-cone tree amplitudes,
and the proof of their Lorentz covariance.
Our main motivation  is the issue of unitarity of non-critical strings
for which the  approach
based solely on the physical degrees of freedom is especially well suited.
Although expected, it
is a priori not obvious that in non-critical dimensions the light-cone construction
is equivalent to the covariant one. Our results
suggest that this is the case. If so, the
light-cone approach would provide a useful complementary technique  for
analyzing non--critical strings.
It also yields
the physical interpretation of  string interaction in terms of
subsequent joining and splitting processes.

Our considerations are based on the following assumptions concerning
the correlators of the Liouville theory on punctured spheres:
\begin{enumerate}
\item
the spectrum of the free external states is described by the Verma module
with the central charge $1< c^{\rm \scriptscriptstyle L}=1+48\beta <25$
and the highest weight $(2\beta,2\beta)$;
\item
the standard form of the conformal anomaly and its relation to the central charge is preserved;
\item
the conformal Ward identities  hold for the punctures and for the energy-momentum tensor itself.
\end{enumerate}
Save for the first assumption (determined by the spectrum of free string models under consideration)
these are the general properties of the Liouville theory in all hitherto approaches.
Our main result is that the assumptions stated above are necessary and sufficient
conditions both for the construction of the light-cone amplitudes and for their Lorentz covariance.

The paper is organized as follows. In Section 2 the free massive
string model in the light-cone formulation is defined and the
light-cone local fields in each sector are introduced. In Section
3 following the idea of joining--splitting interactions we
construct the tree string amplitudes. In Subsection 3.2 we discuss
all the properties necessary for the construction of the amplitude
in the Liouville sector. On this level the conformal Ward
identities are indispensable  for the state--puncture correspondence.
In Subsection 3.4 the final form of the light-cone amplitude is
calculated. In particular an explicitly Lorentz covariant form of
the tachionic tree amplitude is derived. It coincides with the corresponding
covariant amplitude. This indicates that both formulations are indeed
equivalent also in the non-critical
dimensions\footnote{A formal proof is more complicated that in the critical string case \cite{gw}
where one
can use the factorization argument to avoid technically involved calculations for
excited states.}.
In Section 4 we present
the proof of the Lorentz covariance for arbitrary tree amplitude.
It turns out that the properties of the Liouville sector
introduced in Subsection 3.2. are sufficient to complete the
proof. Finally, in Section 5, we discuss the issue of unitarity and
derive some conclusions.

We attached three appendices containing detailed derivations of the
results used in the main text.

\section{Free string}
\setcounter{equation}{0}

\subsection{Light-cone formulation}

In this subsection we remind the light-cone formulation of closed
bosonic massive string propagating in the $d$ dimensional
Minkowski target space ($1<d<25$) \cite{dahaja98}.

For a given choice of a light-cone basis one  construct the quantum theory
as a representation of the algebra of zero modes
$$
[P^{i},x^j] \;  =\;  -  i\delta^{ij}\ ,
\hskip 5mm
[P^+,x^-] \;  =\;  i\ ,
$$
along with the  algebra of the {\it transverse}
$$
\begin{array}{rlllll}
{[a_m^i,a_n^j]} &=&  m\delta^{ij}\delta_{m,-n}\ ,&\\
{[\widetilde{a}_m^i,\widetilde{a}_n^j]}&=&
m\delta^{ij}\delta_{m,-n} \ , &\;\;\;m,n\in \mathbb{Z}
\setminus\{0\}
\end{array}
$$
the {\it Liouville longitudinal}
$$
\begin{array}{rlllll}
{[c_m,c_n]} &=&  m\delta_{m,-n}\ ,
&\\
{[\widetilde{c}_m,\widetilde{c}_n]}
 &=&
 m\delta_{m,-n} \ ,
 &\;\;\; m,n\in \mathbb{Z} \setminus\{0\}
\end{array}
$$
and the {\it Brower longitudinal}
$$
\begin{array}{rll}
{[L^{\rm \scriptscriptstyle B}_m,L^{\rm \scriptscriptstyle B}_n]}
&=& (m-n) L^{\rm \scriptscriptstyle B}_{m+n} + {c^{\rm
\scriptscriptstyle B}\over 12} (m^3 - m) \delta_{m,-n}\ ,
\\ [6pt]
{[\widetilde{L}^{\rm \scriptscriptstyle B}_m, \widetilde{L}^{\rm
\scriptscriptstyle B}_n]} &=& (m-n) \widetilde{L}^{\rm
\scriptscriptstyle B}_{m+n} + {c^{\rm \scriptscriptstyle B}\over
12} (m^3 - m) \delta_{m,-n}\ ,
\end{array}
$$
excitations.   All other commutators vanish and the standard conjugation properties are
assumed.
Although heterotic constructions are possible
we shall consider only the models with the same central charge in the right and the left
algebras of Brower longitudinal excitations. This is also the case which
arises in the covariant quantization of closed  massive string \cite{haja}.

The algebra of non-zero modes is by construction isomorphic to the
(diagonalized) algebra of the DDF operators of the covariant
approach \cite{haja}.   In particular, the algebra of
$L^{\rm \scriptscriptstyle B}_m$ corresponds to the (left) Virasoro
algebra of the shifted Brower longitudinal DDF operators.
There is no zero mode in this sector
and the space of states  has the structure of
tensor product of the left and the right
Verma modules.
In the following we shall restrict ourselves to the discrete series of
the unitary Verma modules ${\cal V}_m(p,q)$ with the central charge
\cite{frqish,gokeol}
$$
 c^{\rm \scriptscriptstyle B}=
c_m\equiv 1 - {6\over m(m+1)} \ ,\;\;\;m=2,3,\ldots \ ,
$$
and with  the highest weight
$$
h_m(p,q)  = {((m+1)p -m q)^2 - 1 \over 4m(m+1)}\ , \hskip 5mm
1\leqslant p\leqslant m-1\ ,\;\;\; 1\leqslant q\leqslant p\ .
$$
Any pair of the allowed left and right highest weights leads to
a consistent free string model. Slightly more general models can be constructed as
direct sums of pairs of Verma modules of the
same central charge and
with different highest weights. Guided  by the experience with the critical string we
 assume that the space ${\cal V}^{\rm \scriptscriptstyle B}$ of Brower
longitudinal  excitation is identical with
the space of states of a unitary minimal model of
 2-dim conformal field theory.
In the  case  of the diagonal ($A$-type)  minimal models one gets \cite{caitzu87}
$$
{\cal V}^{\rm \scriptscriptstyle B}=
\bigoplus\limits_{p,q}{\cal V}_m(p,q)\otimes \widetilde{{\cal V}}_m(p,q)
\ .
$$
The space of states in the transverse sector is given by
$$
 {\cal H}^{\rm \scriptscriptstyle T}=
\int
 d^{d-2}\overline{p}\;\;
 {\cal F}(\overline p)
\ ,
$$
where $\overline{p}= (p^1,\ldots,p^{d-2})$, and
${\cal F}(\overline{p})$ denotes the Fock space
generated by the left and right transverse  excitations
out of the unique ground state $|\overline{p}\;\rangle$
satisfying
$$
P^i\,|\overline{p}\;\rangle = p^i|\overline{p}\;\rangle
\ .
$$
The total space of states in the light-cone formulation is then defined by
$$
{\cal H}=
\int {dp_+  }\; | p^+\rangle\otimes
 {\cal H}^{\rm \scriptscriptstyle T}\otimes
{\cal V}^{\rm \scriptscriptstyle L}\otimes{\cal V}^{\rm \scriptscriptstyle B}
\ ,
$$
where $P^+ | p^+\rangle= p^+| p^+\rangle$ and
${\cal V}^{\rm \scriptscriptstyle L}$ is the Fock space
generated by the left and right  Liouville excitations
out of the unique vacuum state $|0\rangle^{\rm \scriptscriptstyle L}$.

In order to construct a unitary realization of the Poincar\'{e}
algebra on ${\cal H}$ we introduce
\begin{eqnarray}
L_{n}^{\rm \scriptscriptstyle T} &=&
 {\textstyle\frac{1}{2}}\sum_{k= -\infty}^{+\infty}
 \sum_{i=1}^{d-2}
 :\!a_{-k}^ia_{n+k}^i\!:
 \ , \nonumber \\
 L_{n}^{\rm \scriptscriptstyle L} &=&
 {\textstyle\frac{1}{2}}\sum_{k\neq 0,-n}\!
 :\!c_{-k}c_{n+k}\!: \;+\;2i\sqrt{\beta} n c_n \;+\;
 2\beta\delta_{n,0}
 \ . \label{LL}
\end{eqnarray}
The operators
$
L_n\equiv
L_{n}^{\rm \scriptscriptstyle T} +
L_{n}^{\rm \scriptscriptstyle L} +
L_{n}^{\rm \scriptscriptstyle B}
$
and their left counterparts $\widetilde{L}_n$
form two commuting Virasoro algebras with the central charge
$
c = d + 48\beta   -
{\textstyle{6\over m(m+1)}}
$.
In this construction we have used the zero modes
 $a^i_0= \widetilde{a}^i_0 ={1\over 2\sqrt{\alpha}} P^i$
with the  dimensionful parameter $\alpha$ related to the conventional
Regge slope $\alpha'$  by $\alpha = {1\over 2 \alpha'}$.
The  algebra of  generators:
\begin{eqnarray}
{P}^- &=&
{\frac {2\alpha}{ P^+}} (L_0 +
\widetilde{L}_0 - 2) \ ,\;\;\;P^+\ ,\;\;\; P^i\ ,
\nonumber\\
{M}^{ij}_{\rm lc} & = & {P}^i{x}^j-{P}^j{x}^i + i\sum_{n\geqslant
1} {\frac{1}{n}} \left (a_{-n}^ia_{n}^j - a_{-n}^ja_{n}^i +
\widetilde a_{-n}^i\widetilde a_{n}^j - \widetilde
a_{-n}^j\widetilde a_{n}^i\right) \ ,
\nonumber \\
{M}^{i+}_{\rm lc} & = & {P}^+ {x}^i \ ,
\label{poincare}\\
{M}^{+-}_{\rm lc} &=& {\textstyle\frac{1}{2}}({P}^+{x}^-+x^-P^+) \ ,
\nonumber \\
{M}^{i-}_{\rm lc} &=& {\textstyle\frac{1}{2}}( {x}^i P^-
 + P^- {x}^i ) - {P}^i{x}^- -
 i{\frac{ 2\sqrt{\alpha}}{P^+}} \sum_{n\geqslant 1} {\frac{1}{n}}
\left( a_{-n}^i L_n
- L_{-n} a_n^i
+ \widetilde a_{-n}^i \widetilde L_n
- \widetilde L_{-n} \widetilde a_n^i
\right)\ ,  \nonumber
\end{eqnarray}
 closes to the Lie algebra of Poincar\'{e} group if and only if
the parameter $\beta$ entering the definition of
 $ L_n^{\rm \scriptscriptstyle L} $ satisfies \cite{daja00}:
$$
\beta =
 \;{24-d\over 48} + {1\over 8m(m+1)}
 \ .
$$
This representation induces a unitary representation on the
subspace ${\cal H}^{\rm \scriptscriptstyle ph}\subset {\cal H}$ of
physical closed string states  defined by the condition
$$
(L_0 - \widetilde{L}_0)|\Psi\rangle =0 \ .
$$
In the case of the diagonal minimal model in the Brower
longitudinal sector the on-mass-shell condition takes the form
$$
M^2 = 4\alpha \left( 2N + {((m+1)p - m q )^2 \over 2m(m+1)}
 - {d\over 12}
\right) \ ,\;\;\;N=0,1,2,\dots \ .
$$

\subsection{Light-cone fields in the transverse sector}

Following standard light-cone formulation
\cite{Mandelstam73,Mandelstam74,Mandelstam86} we choose the  world sheet
 parameterization $(\sigma,\tau)$ where $\tau$ is related to the target space time by
$\tau = 2\sqrt{\alpha}x^+$ and $\sigma$ is in
the range\footnote{The subscript $r$ will serve in the following
to distinguish between various incoming and outgoing strings; we
use it here to avoid confusing the range of $\sigma$
with the parameter  $\alpha = {1\over 2 \alpha'}$.}
$$
0 \leqslant \sigma < 2\pi\left|\alpha_r\right|\ ,\;\;\; \alpha_r
\;=\;\frac{p_r^{+}}{\sqrt\alpha}\ .
$$
The string fluctuations in the transverse sector are  described
by the fields
$$
X^i(\sigma,\tau)  = x^i + \frac{P^i }{2 \alpha}
\frac{\tau}{\alpha_r} + \frac{i}{2\sqrt\alpha}
\sum\limits_{n\neq 0} \frac{1}{n}{\rm e}^{-\frac{in\tau}{\alpha_r}}
\left(a_n^i{\rm e}^{-\frac{in\sigma}{\alpha_r}}
+ \widetilde a_n^i{\rm e}^{\frac{in\sigma}{\alpha_r}}\right)
$$
satisfying the Heisenberg equation of motion
$$
2\sqrt\alpha {\partial \over \partial \tau} X^i(\sigma,\tau)
= i [ P^-,X^i(\sigma,\tau)] \ .
$$
After Wick rotation $\tau \to -i\tau$ the  fields $X^i$ can be
decomposed into the holomorphic and antiholomorphic components in
the complex variable $\rho = \tau + i\sigma$:
$$
\begin{array}{rll}
X^i(-i\tau,\sigma) &=&\frac{1}{2\sqrt\alpha}\left(\chi^i(\rho) +
\widetilde\chi^i(\bar\rho)\right)\ ,\\
\chi^i(\rho)&= & q^i_0 -ia^i_0\frac{\rho}{\alpha_r} +
i\sum\limits_{n\neq 0}\frac{a^i_n}{n}{\rm e}^{-\frac{n\rho}{\alpha_r}}\ ,\\
\widetilde\chi^i(\bar\rho) &= &
\widetilde{q}^i_0 -i\widetilde a^i_0\frac{\bar\rho}{\alpha_r} +
i\sum\limits_{n\neq 0}\frac{\widetilde a^i_n}{n}{\rm e}^{-\frac{n\bar\rho}{\alpha_r}}\ ,
\end{array}
$$
where the left and the right zero modes satisfy:
$$
\begin{array}{rlll}
[a_0^i, q_0^j] &=&- i\delta^{ij}\ ,\;\;\;
[\widetilde{a}_0^i,\widetilde{q}_0^j] \;=\; - i\delta^{ij}\ ,\\
P^i &=& \sqrt\alpha (a_0^i+\widetilde{a}_0^i) \ ,\\
x^i &=& {1\over 2 \sqrt\alpha}(q_0^i+\widetilde{q}_0^i)\ ,
\end{array}
$$
and the restriction to the  subspace
on which $a_0^i=\widetilde{a}_0^i$ is assumed.
Another local field of the transverse sector is the energy-momentum tensor.
Within the Euclidean framework its holomorphic
component  is given by
$$
{T^{\rm \scriptscriptstyle T}(\rho)}
 =
\frac{1}{\alpha_r^2}\sum\limits_{n}\left(L_n^{\rm \scriptscriptstyle T} -
\frac{d}{24}\delta_{n,0}\right)
{\rm e}^{-\frac{n\rho}{\alpha_r}} \ .
$$
In order to avoid unnecessary repetition we omit here and in the
following  the  formulae for the right (antiholomorphic) sector.
The generator $Q_{\epsilon,\tilde\epsilon}$ of the the
infinitesimal conformal transformation $ \rho \to \rho +
\epsilon(\rho)\;,\;\bar\rho\to \bar\rho+\tilde\epsilon(\bar\rho) $
reads
\begin{equation}
\label{generator} Q^{\rm\scriptscriptstyle
T}_{\epsilon,\tilde\epsilon} \equiv {1\over 2\pi i}\int\limits
d\rho \,\epsilon(\rho) T^{\scriptscriptstyle T}(\rho)
 +
{1\over 2\pi i}\int\limits d\bar\rho \,
\widetilde\epsilon(\bar\rho) \widetilde T^{\scriptscriptstyle
T}(\bar\rho)\ ,
\end{equation}
and the transformation rules  for the transverse fields
and  states
take the form
\begin{eqnarray}
\label{chiward}
\delta_{\epsilon} \chi^i(\rho)
&\equiv& - \, \epsilon(\rho)\partial_\rho   \chi^i(\rho)
 = - \,[Q^{\rm\scriptscriptstyle T}_{\epsilon,\tilde\epsilon},\chi^i(\rho) ]   \ ,\\
\label{ttward}
 \delta_{\epsilon}T^{\rm\scriptscriptstyle T}(\rho)
&=& - \,[Q^{\rm\scriptscriptstyle T}_{\epsilon,\tilde\epsilon},
T^{\rm\scriptscriptstyle T}(\rho)  ]   \ , \\
\label{tstates}
\delta_{\epsilon,\tilde\epsilon}\,|\Psi\rangle &=&
-Q^{\rm\scriptscriptstyle T}_{\epsilon,\tilde\epsilon}\,|\Psi\rangle \ .
\end{eqnarray}
For arbitrary conformal mapping $\rho \to \rho(z)$ the integrated version of
(\ref{ttward}) yields
\begin{equation}
\label{ttlow}
T^{\rm \scriptscriptstyle T}(\rho) \to
T'{}^{\rm \scriptscriptstyle T}(z) =
\left(\frac{d\rho(z)}{dz}\right)^2 T^{\rm \scriptscriptstyle T}\big(\rho(z)\big) +
 \frac{c^{\rm \scriptscriptstyle T}}{12}S(\rho,z)
\end{equation}
where the Schwartz derivative is defined as
$$
S(\rho,z) = \frac{\rho^{(3)}(z)}{\rho'(z)} -
\frac32 \left(\frac{\rho''(z)}{\rho'(z)}\right)^2\ .
$$
The transformation law (\ref{ttlow}) is satisfied by the free
field representation of the energy--mo\-men\-tum tensor
$$
T^{\rm\scriptscriptstyle T}(\rho) =
-\frac12:\!\partial_\rho\chi^i(\rho)\partial_\rho\chi_i(\rho)\!:
 - \frac{d}{24\alpha_r^2} \ .
$$
By means of the conformal map
$
z \rightarrow \rho (z) =  \alpha_r \ln \rho
$
one can proceed to the complex plane where
 $$
 \begin{array}{rll}
T^{\rm \scriptscriptstyle T}(z) &=& \sum\limits_{n\in \mathbb{Z}}
L^{\rm \scriptscriptstyle T}_n z^{-n-2} \ ,\\ [8pt] \chi^i(z) & =
& \chi^i_0 -ia^i_0\ln z + i\sum\limits_{n\neq 0}\frac{a^i_n}{n}
z^{-n}\ .
\end{array}
$$
According to the standard operator--state correspondence for each
state $|\Psi\rangle \in {\cal H}^{\rm \scriptscriptstyle T}$ there
exists a unique local vertex operator $V_\Psi(z,\bar z )$ such
that
$$
| \Psi\rangle =
\lim\limits_{z,\bar z \to 0}V_{\Psi}(z,\bar z)|\overline{0}\;\rangle \ ,
$$
and $|\,\overline{0}\,\rangle $ is the unique $P
SL(2,\mathbb{C})$-invariant vacuum state in ${\cal H}^{\rm
\scriptscriptstyle T}$. In particular the vertex operators
$V_{\overline{p}}(z,\bar z)$ corresponding to the ground states $|
\overline{p}\;\rangle \in {\cal H}^{\rm \scriptscriptstyle T}$ are
given by
$$
V_{\overline{p}}(z,\bar z)\; \equiv \;\ :{\rm e}^{{i\over
2\sqrt{\alpha}}p^i (\chi^i(z) +\widetilde\chi^i (\bar z))}\!:\ .
$$
One possible way to construct the vertex operator corresponding to
the excited state is to express the creation operators in terms of
the contour integrals of the local fields $X^i(w,\bar w)$. Another
way, which can be easily  generalized to the longitudinal sector,
is to consider  the transverse sector as a tensor product ${\cal
H}^{\rm\scriptscriptstyle T} =\prod\limits_{i=1}^{d-2}{\cal H}^i$
of $d-2$ copies of the $c=1$ scalar CFT. In each ${\cal H}^i$ one
has the basis consisting of the vectors
 $$
\left.\left|\{\mbox{\boldmath $k$}\}, \{\overline{\mbox{\boldmath
$k$}}\} ,  p^i\right.\right\rangle \; = \; \widetilde
L^i_{r,-\bar k_1}\ldots\widetilde L^i_{r,-\bar k_m} L^i_{r,-
k_1}\ldots L^i_{r,- k_n}| p^i\,\rangle\ ,
$$
generated by the modes $L^i_{r,- k},$ $\widetilde L^i_{r,-\bar
k},$  of the $i^{\rm th}$ components $T^i(\rho), \widetilde
T^i(\bar \rho)$ of the ``transverse'' energy--momentum tensor
$T^{\rm\scriptscriptstyle T}(\rho_r), \widetilde
T^{\rm\scriptscriptstyle T}(\bar \rho_r)$. The vertex operators
corresponding to these states can be constructed as
\begin{eqnarray*}
V^i_{|\{\mbox{\boldmath $\scriptscriptstyle k$}\},
\{\overline{\mbox{\boldmath $\scriptscriptstyle k$}}\}
,p^i\rangle}(z_r,\bar z_r) & = & \oint\limits_{\bar
z_r}\frac{d\bar w_1}{2\pi i} \frac{\widetilde T^i(\bar
w_1)}{(\bar w_1-\bar z_r)^{\bar k_1 -1}} \ldots
\oint\limits_{\bar z_r}\frac{d\bar w_n}{2\pi i} \frac{\widetilde
T^i(\bar w_m)}{(\bar
w_m-\bar z_r)^{\bar k_n -1}} \\
&& \hskip 5mm \oint\limits_{z_r}\frac{dw_1}{2\pi i}
\frac{T^i(w_1)}{(w_1-z_r)^{k_1 -1}} \ldots
\oint\limits_{z_r}\frac{dw_m}{2\pi i}
\frac{T^i(w_m)}{(w_m-z_r)^{k_m -1}} \; :{\rm e}^{ip^iX^i(z_r,\bar
z_r)}\!: \ ,
\end{eqnarray*}
where the integration contours are nested radially around $z_r,~$
$|w_1-z_r| > |w_2 -z_r| > \ldots > |w_m - z_r|,~$ $|\bar w_1-\bar
z_r| > |\bar w_2 -\bar z_r| > \ldots > |\bar w_m -\bar z_r|.$

\subsection{Light-cone fields in the longitudinal sector}

The CFT interpretation of the Brower longitudinal sector is essentially the
same as in the transverse one except there are no counterparts to the fields
 $\chi^i, \widetilde\chi^i,$
which reflects the fact that there is no classical description of  this sector in terms of
action functional.
 The energy--momentum tensor and the transformations rules on the cylinder are
 given  by the standard expressions:
\begin{eqnarray}
\label{btensor} {T^{\rm \scriptscriptstyle B}(\rho)} &=&
\frac{1}{\alpha_r^2}\sum\limits_{n}\left(L_n^{\rm
\scriptscriptstyle B} - \frac{c^{\rm \scriptscriptstyle
B}}{24}\delta_{n,0}\right) {\rm e}^{-n{\rho\over \alpha_r}}\ ,
\\
\label{btward}
\delta_{\epsilon}T^{\rm\scriptscriptstyle B}(\rho)
&=&- \,[Q^{\rm\scriptscriptstyle B}_{\epsilon,\tilde\epsilon},
T^{\rm\scriptscriptstyle B}(\rho)  ]   \ ,\\
\label{bstates}
\delta_{\epsilon,\tilde\epsilon}|\Psi\rangle &=&
-Q^{\rm\scriptscriptstyle B}_{\epsilon,\tilde\epsilon}|\Psi\rangle \ .
\end{eqnarray}
Due to the assumed minimal model structure there exists the unique
$SL(2,\mathbb{C})$-invariant vacuum state
$|0,0\rangle^{\rm\scriptscriptstyle B}\in {\cal
V}^{\rm\scriptscriptstyle B}$ and the standard operator--state
correspondence holds. Proceeding to the complex plane by the map
$\rho=\alpha_r\ln z$ one gets:
\begin{eqnarray}
\label{btensorplane}
{T^{\rm \scriptscriptstyle B}(z)}
&=&
\sum\limits_{n}L_n^{\rm \scriptscriptstyle B}z^{-n-2}
\ ,
\\
\label{btwardplane}
\delta_{\epsilon}T^{\rm\scriptscriptstyle B}(z)
&=&- {1\over 2\pi i}\oint dw\, \epsilon(w) T^{\rm\scriptscriptstyle B}(w)
T^{\rm\scriptscriptstyle B}(z)   \ ,\\
\label{bstatesplane}
\delta_{\epsilon,\tilde\epsilon}V_\Psi(z,\bar z) &=&
-{1\over 2\pi i}\oint \,d w\,
 \epsilon(w) T^{\scriptscriptstyle B}(w)V_\Psi(z,\bar z) \\
 &&
- {1\over 2\pi i}\oint \,d \bar w\, \widetilde\epsilon(\bar w)
\widetilde T^{\scriptscriptstyle B}(\bar w)V_\Psi(z,\bar z) \ .
\nonumber
\end{eqnarray}
Let us note that in contrast to the transverse sector where the
free field representation of the energy--momentum tensor is
available, the l.h.s. of (\ref{btwardplane}) can be independently
defined only by the transformation properties of the correlation
functions involving $T^{\rm\scriptscriptstyle B}(z)$.
\bigskip

The space of states in the Liouville sector carries
representations of two different Virasoro algebras  with the
central charges $c= 1 + 48\beta$ and $c = 1$. This leads to two
different interpretations of this sector as a conformal field
theory.

Within the   {\em scalar interpretation} the space of states in
the Liouville sector can be seen as the restriction of the space
of states of a single scalar field in two dimensions to the
subspace of states with zero momentum. In order to make the
standard field theoretic techniques available one can extend the
space of states to that of a single scalar field and then define
the subspace of physical states by imposing an extra constraint of
vanishing momenta. As in the case of the transverse sector it is
convenient to introduce the algebra of left and right zero modes
$$
[c_0, \phi_0] =- i\ ,\;\;\;
[\widetilde{c}_0,\widetilde\phi_0] \;=\; - i\ ,
$$
restricted to the subspace annihilated by $c_0 -\widetilde{c}_0$.
The subspace of physical states is then determined by the
condition $(c_0 +\widetilde{c}_0)|\Psi\rangle = 0$. The
Poincar\'{e} representation on the extended space is given by the
formulae (\ref{poincare}) with the  Virasoro  generators
$L_n^{\scriptscriptstyle L}$ (\ref{LL}) modified by the
appropriate zero mode contribution. In the extended space one can
define the local Liouville field
\begin{equation}
\label{liouville}
\phi(\rho)= \phi_0 -ic_0{\rho\over \alpha_r}+
i\sum\limits_{n\neq 0}\frac{c_n}{n}{\rm e}^{-n{\rho\over \alpha_r}}\ .
\end{equation}
The energy--momentum tensor and the transformation rules are given
as in the Liouville interpretation by the formulae (\ref{btensor}
-- \ref{bstatesplane}). The crucial feature of the scalar
interpretation is that the transformation rule for the Liouville
field,
\begin{equation}
\label{ltrans}
\delta_{\epsilon} \phi(\rho)
 = - \,[Q^{\rm\scriptscriptstyle L}_{\epsilon,\tilde\epsilon},\phi(\rho) ]
 \;=\; -\,\epsilon(\rho)\partial_\rho \phi(\rho)   \;-\,
2\sqrt\beta  \partial_\rho \epsilon(\rho)\ ,
\end{equation}
does not coincide with the change of $\phi(\rho)$ under
infinitesimal conformal transformation of its argument  $\rho \to
\rho + \epsilon(\rho)$. The same discrepancy shows up in the
transformation rule of the energy--momentum tensor which can  be
written as
\begin{equation}
\label{ltensorfields}
T^{\rm\scriptscriptstyle L}(\rho) =
-\frac12:\!\partial_\rho\phi(\rho)\partial_\rho\phi(\rho)\!:
+2\sqrt\beta\partial^2_\rho\phi(\rho) - \frac{1}{24\alpha_r^2} \ .
\end{equation}
It satisfies formulae (\ref{btward}), (\ref{btwardplane}) only
when the non-homogeneous transformation law for the Liouville
field is assumed. The drawback of the scalar interpretation  is
that the local field defined by (\ref{liouville}) does not provide
any working model for the transformation law (\ref{ltrans}). One
can show that just this feature is responsible for the failure of
constructing Lorentz covariant joining--splitting interactions
within the scalar approach. For this reason we shall not discuss
this interpretation  in this paper any more.

According to the {\it Liouville interpretation} the space of
states  is a tensor product of single left and single right Verma
modules with the same central charge $c^{\rm \scriptscriptstyle L}
=\bar c^{\rm \scriptscriptstyle L} = 1 + 48\beta$ and the same
non-zero highest weight $h=\bar h = 2\beta$. The construction of
energy--momentum tensor and the form of transformation rules are
essentially the same as in the Brower longitudinal sector and are
given by exact counterparts of the formulae (\ref{btensor} --
\ref{bstatesplane}). The crucial difference is that the only
ground state $|\,0\,\rangle^{\rm \scriptscriptstyle L}$ in this
sector is not $PSL(2,\mathbb{C})$-invariant,
$$
L^{\rm \scriptscriptstyle L}_0|\,0\,\rangle^{\rm
\scriptscriptstyle L}= 2\beta\,|\,0\,\rangle^{\rm
\scriptscriptstyle L} \ ,\;\;\;\;\;\; \widetilde L^{\rm
\scriptscriptstyle L}_0|\,0\,\rangle^{\rm \scriptscriptstyle L}=
2\beta\,|\,0\,\rangle^{\rm \scriptscriptstyle L} \ ,
$$
and  the energy momentum-tensor acting on $|\,0\,\rangle^{\rm
\scriptscriptstyle L}$ is singular in the limit $z\to 0,$
$$
T^{\rm \scriptscriptstyle L}(z)|\,0\,\rangle^{\rm \scriptscriptstyle L}
= {2\beta \over z^2}|\,0\,\rangle^{\rm \scriptscriptstyle L}
+ {1\over z}L^{\rm \scriptscriptstyle L}_{-1}|\,0\,\rangle^{\rm \scriptscriptstyle L}
+ {\rm regular \; terms} \ .
$$
This singular behavior cannot be mimic by any local operator
placed at the origin of the complex plane which means that the
operator--state correspondence does not hold in this sector.
Instead of the identity operator which corresponds in the standard
CFT to the invariant vacuum, there is a puncture   corresponding
to the non invariant vacuum state $|\,0\,\rangle^{\rm
\scriptscriptstyle L}$ with prescribed  singularity structure of
the energy--momentum tensor at it.
 It should be stressed that
 the Liouville interpretation goes far beyond the standard
 CFT where the existence of a unique invariant vacuum and the operator--state
 correspondence are fundamental assumptions.

\section{Interacting string}

\setcounter{equation}{0}

\subsection{Joining--splitting interactions}
In the light-cone picture, developed by S. Mandelstam
\cite{Mandelstam73,Mandelstam74,Mandelstam86}, string amplitudes
are constructed in terms of light-cone diagrams corresponding to
time ordered sequences of elementary splitting and joining
processes.

The elementary process of joining--splitting can be described in
terms of
 the vertex operator acting between three  Hilbert spaces of free string
states \cite{crge,kaki,hotuco,grsc83}. Assuming the free string
evolution  between acts of interaction one can construct string
amplitudes automatically satisfying  the  (perturbative) unitarity
requirements. Another approach first introduced by Mandelstam
\cite{Mandelstam73,Mandelstam74} is to express the string
amplitudes in terms of path integral over string trajectories with
the world-sheet geometries corresponding  to appropriate
light-cone diagrams. Within this approach the 3-string interaction
vertex is implicitly defined by the requirement that the free CFT
defined on each cylinder-like region of free string propagation
are glued up to a unique CFT on the light-cone diagram. It can be
shown that in the case of the transverse string excitations
described by the 2-dim scalar CFT both formulations are equivalent
\cite{crge,kaki,hikko}.

In this section we shall follow the original Mandelstam approach and construct
the string amplitudes in terms of
conformal field theories on the light-cone diagrams. It allows to avoid
technically involved questions of sewing conformal field theories and makes it
 easier to express the light-cone amplitudes in terms
of CFT on the complex plane. Let us note that in such formulation
the unitarity  in the longitudinal sector is not automatic
and requires additional considerations.

A tree light-cone diagram  $\Sigma$ with $N$ external states is
uniquely characterized by the circumferences $2\pi\alpha_r$ of
external semi-infinite cylinders, the interaction times
$\tau_1\leqslant\tau_2 \leqslant \dots\leqslant \tau_{N-2}$, and
the $N-3$ angles  describing twists of the intermediate cylinders.
These twist angles can be parameterized by the $\sigma_I$
coordinates of all but one interaction points. We assume that all
external states are eigenstates of the momentum operators
$P^+,P^-,P^i$. In the $p^+$ sector  the tree  string light-cone
amplitude is given by the conservation delta function. The
eigenvalues $p^+_r$ of the external states are related to the
geometry of $\Sigma$ by $p^+_r= \sqrt{\alpha}\alpha_r$.  If we
assume the time translation invariance the integration over one of
the interaction times results in the energy conservation delta
function. The contribution of the light-cone diagram $\Sigma$ to
the string amplitude can than be written in the form
\begin{equation}
\label{amplitude}
A_\Sigma =\delta\left(\sum\limits_{r=1}^N p_r^+\right)
\delta\left(\sum\limits_{r=1}^N p_r^-\right)
\int\prod\limits_{I=2}^{N-2}d^2(\rho_I-\rho_1)
    \;W^{\scriptscriptstyle\rm T}
    W^{\scriptscriptstyle\rm B}W^{\scriptscriptstyle\rm L}
\end{equation}
where $\rho_I = \tau_I + i\sigma_I$ describe locations of the interaction points
and the integration domain reflects the time ordering of string interactions.
$W^{\scriptscriptstyle\rm T},W^{\scriptscriptstyle\rm B},W^{\scriptscriptstyle\rm L}$
denote the transverse, the Brower, and the Liouville sector contributions respectively.

\subsection{Transverse sector}
In this subsection we shall briefly discuss the transverse sector
contribution to the tree string amplitude. This material is not
new and was included in order to set notation and to provide an
introduction for the discussion of  the longitudinal sector. The
only novelty is the derivation of the amplitude in terms of
correlation functions of vertex operators in the case of excited
physical states. The transverse sector contribution is defined by
\begin{eqnarray}
 \label{amplit_1}
  W^{\scriptscriptstyle\rm T}
  &= &  \prod\limits_r\int\limits_0^{2\pi\alpha_r}\!d\vartheta_r
  \int\!{\cal D}X^i_r(\sigma_r)\;
  {\rm e}^{-\frac{1}{2\sqrt\alpha}
  (P_r^{-})^{\rm \scriptscriptstyle T}\tau_r}
  {\textstyle {1\over \sqrt{2\pi\alpha_r}}}
  \Psi_r[X^i_r(\sigma_r + \vartheta_r)]
 \\
 &&\hspace{30pt}\times \int_\Sigma\!{\cal D}^{g_{\scriptscriptstyle\rm lc}}X^i(\sigma,\tau)\;
  {\rm e}^{-S[g_{\scriptscriptstyle\rm lc},X^i(\sigma,\tau)]}
  \ ,\nonumber
\end{eqnarray}
where the inner path integral goes over fields satisfying the
Dirichlet boundary conditions
$
X^i(\sigma_r,\tau_r) = X^i_r(\sigma_r)
$.
According to our assumption  the external states are eigenstates of the
transverse part $(P_r^{-})^{\rm \scriptscriptstyle T} $
of the light-cone Hamiltonian,
$$
\begin{array}{l}
(P_r^{-})^{\rm \scriptscriptstyle T}\; =\; {\frac {2\alpha}{ P^+}}
(L^{\rm \scriptscriptstyle T}_0 + \widetilde{L}^{\rm
\scriptscriptstyle T}_0 - {c^{\rm \scriptscriptstyle T}\over 12})
\ ,\\ [8pt] \displaystyle \frac{(P_r^{-})^{\rm \scriptscriptstyle
T}}{2\sqrt\alpha}\ \Psi_r[X^i_r(\sigma_r)] \; = \;
\frac{\lambda_r^{\rm \scriptscriptstyle T}}{\alpha_r}\
\Psi_r[X^i_r(\sigma_r)]\ , \\ [8pt]
\lambda_r^{\rm\scriptscriptstyle T}
\; = \; {\overline{p}_r^2\over
4\alpha} +2n^{\rm \scriptscriptstyle T}_r - {c^{\rm
\scriptscriptstyle T}\over 12}\ ,
\end{array}
$$
with the same left $n^{\rm \scriptscriptstyle T}_r$ and  right
$\widetilde n^{\rm \scriptscriptstyle T}_r$ occupation numbers. Then
one  can rewrite (\ref{amplit_1}) in the form
\begin{eqnarray}
 \label{amplit_2}
  W^{\scriptscriptstyle\rm T}&=&
\prod\limits_r \sqrt{2\pi\alpha_r}
{\rm e}^{-\frac{\lambda_r^{\rm\scriptscriptstyle T}}{\alpha_r}\tau_r}
  \int\!{\cal D}X^i_r(\sigma_r)\;\Psi_r[X^i_r(\sigma_r)]
  \\
&&\hspace{80pt}\times  \int_\Sigma\!{\cal D}^{g_{\scriptscriptstyle\rm lc}}X^i(\sigma,\tau)\;
  {\rm e}^{-S[g_{\scriptscriptstyle\rm lc},X^i(\sigma,\tau)]}
  \ .\nonumber
\end{eqnarray}
In order  to express the light cone amplitude in terms of the
conformal field theory on the complex plane we shall use the
holomorphic Mandelstam map
\begin{equation}
\label{Mmap}
 \rho(z) = \sum_r\alpha_r\ln(z-z_r)\ , \hskip 1cm
\sum_r\alpha_r = 0\ .
\end{equation}
At the first step of the calculations one can regard this
 map merely as a change of parameterization.
Let $M_\tau$ be the pre-image  of the light-cone diagram $\Sigma$
with long but finite legs. The inner path integral in
(\ref{amplit_2}) determines on $M_\tau$ the theory of $d$ scalar
fields defined with respect to the pull-back $\rho^*g_{\rm lc}$
of the flat light-cone metric $g_{\rm lc}$. The metric
$\rho^*g_{\rm lc}$  is related to the standard metric $g_{\rm pl}$
on the complex plane by the  conformal factor $\sigma(z),$
$$
\rho^*g_{\rm lc}(z) = {\rm e}^{\sigma(z)}g_{\rm pl}(z)\ ,
 \hskip 1cm
\sigma(z) = \ln \left| {\partial \rho \over
\partial z}\right|^2\ .
$$
It is singular at the locations $Z_I$ of the interaction points and also in the limit
$|z|\to \infty$ (Appendix A.1).

At the second step one uses the conformal properties of the theory
to change the metric from ${\rm e}^{\sigma}g_{\rm pl}$ to $g_{\rm
pl}$. The response of the path integral representation of the
amplitude  to this change is given by the conformal anomaly
\cite{poly81,oalvarez},
\begin{equation}
\label{anomaly_1}
 \int_{M_\tau}\!\!{\cal D}^{{\rm e}^{\sigma}g_{\rm pl}}X(z,\bar z)
 {\rm e}^{-S[{\rm e}^{\sigma}g_{\rm pl},X(z,\bar z)]}
 \; = \;
 ({\cal A}_{M_\tau})^{c^{\rm \scriptscriptstyle T}} \int_{M_\tau}\!\!
 {\cal D}^{g_{\rm pl}}X(z,\bar z) {\rm e}^{-S[g_{\rm pl},X(z,\bar
 z)]}\ ,
\end{equation}
where the integral on the r.h.s. goes over the fields
satisfying boundary conditions
\begin{equation}
\label{bcond_1}
X^i\left(\rho^{-1}(\tau_r + i\sigma_r), \bar\rho^{-1}(\tau_r -
i\sigma_r)\right)  \; = \; X^i_r(\sigma_r)\ .
\end{equation}
In the limit  ${\tau_r\over \alpha_r } \to \infty,\ r=1,\dots,N,$
the anomaly factor takes the form (Appendix A.4)
\begin{equation}
\label{anomaly} {\cal A}_{M_\tau} =[\epsilon]^{{1\over 12}}
 A^{\frac14}
[\alpha]^{-{1\over 8}} [P]^{-{1\over 12 }} \ ,
\end{equation}
where the abbreviated notations (\ref{A}), (\ref{notation}) are
used. Let us stress that since the standard metric $g_{\rm pl}$ is
regular at the interaction points so is the energy--momentum
tensor of the transformed theory.

At the  third step of our calculation we shall express the wave
functionals $\Psi_r[X^i(\sigma_r)]$ through the vertex operators
of the ``transverse'' CFT. Following the standard path integral derivation of
the operator-states correspondence (Appendix B) one gets
$$
\Psi_r[X^i_r(\sigma_r)]  \; = \; {\rm
e}^{(\lambda_r^{\rm\scriptscriptstyle T}-
\frac{c^{\rm\scriptscriptstyle T}}{12})
\frac{\tau_r}{\alpha_r}}
 \int_{D_r}\!\!{\cal D}^{g_{\scriptscriptstyle\rm pl}}
 X^i(z,\bar z)\;
  {\rm e}^{-S[g_{\scriptscriptstyle\rm pl}, X^i(z,\bar z)]}
  \ V_r^{\rm\scriptscriptstyle T}(z_r,\bar z_r)\ ,
$$
where the integration goes  over the fields defined on
the disc
$$
D_{\eta_r}\; = \; \left\{z \in \mathbb{C}\; : \: |z-z_r| \leqslant
\eta_r = {\rm e}^{\frac{\tau_r}{\alpha_r}}\right\}
$$
and satisfying the boundary conditions
$$
X^i\left(z_r +{\rm e}^{\frac{\tau_r+ i\sigma_r}{\alpha_r}}, \bar
z_r +{\rm e}^{\frac{\tau_r- i\sigma_r}{\alpha_r}}\right) =
X^i_r(\sigma_r)\ ,
$$
while $V_r^{\rm\scriptscriptstyle T}(z_r,\bar z_r)$ is the vertex
operator corresponding  to the
state $|\,r\,\rangle^{\rm\scriptscriptstyle T}.$

The complement of $M_\tau$
 consists of disjoint sets $D_{\tau_r}$ around each point $z_r$.
For every $r$ we define a non-singular, invertible conformal transformation
\begin{equation}
\label{conf_map}
 w_r(z) \; = \; \rho_r^{-1}\left(\rho(z)\right) \; =
\; z_r + (z-z_r)\prod\limits_{s,s\neq
r}(z-z_s)^{\frac{\alpha_s}{\alpha_r}} \ ,
\end{equation}
from $D_{\tau_r}$ into $D_{\eta_r}$, with $z_r$ being a fixed point.
Using this map one can pull all the objects back
from $D_{\eta_r}$ to $D_{\tau_r}$.
This yields the representation of
 the $r^{\rm th}$ string wave functional
\begin{eqnarray}
\label{amplit_8} \Psi_r[X^i_r(\sigma_r)]  & = & {\rm
e}^{\left(\lambda_r^{\rm\scriptscriptstyle T}-
\frac{c^{\rm\scriptscriptstyle T}}{12}\right)
\frac{\tau_r}{\alpha_r}} \left({\cal A}_{D_{\tau_r}}
\right)^{c^{\rm\scriptscriptstyle T}}
\\
 &&\hspace{30pt}\times
 \int_{D_{\tau_r}}\hspace*{-4mm}{\cal D}^{g_{\scriptscriptstyle\rm pl}}
  X^i(z,\bar z)\;
  {\rm e}^{-S[g_{\scriptscriptstyle\rm pl}, X^i(z,\bar z)]}
  \ w_r^*V_r^{\rm\scriptscriptstyle T}(z_r,\bar
  z_r)\ ,\nonumber
\end{eqnarray}
where ${\cal A}_{D_{\tau_r}}$ denotes the conformal anomaly of the
map (\ref{conf_map}) and the integration goes over the fields $X^i(z,\bar
z)$ on $D_{\tau_r}$ satisfying boundary condition
\begin{equation}
\label{bcond_2}
 X^i\left(\rho^{-1}(\tau_r + i\sigma_r),
\bar\rho^{-1}(\tau_r - i\sigma_r)\right)\; = \; X^i_r(\sigma_r)\ .
\end{equation}
$w_r^*V_r^{\rm\scriptscriptstyle T}(z_r,\bar z_r)$ in formula (\ref{amplit_8})
denotes the pull-back of the vertex operator $V_r^{\rm\scriptscriptstyle
T}(z_r,\bar z_r)$ by the map (\ref{conf_map}).
In the simplest case of the primary field corresponding to the ground state
$|\vec p\,\rangle$ we have
\begin{eqnarray}
\label{w*vertex}
w_r^*V_{\vec p}^{\rm\scriptscriptstyle T}(z_r,\bar z_r)
& = &
 \left(\frac{dw_r}{dz}(z_r) \right)^{-h_r}
 \left(\frac{d\bar w_r}{d\bar z}(\bar z_r) \right)^{-\widetilde h_r}
 V_{\vec p} ^{\rm\scriptscriptstyle T}(z_r,\bar z_r) \\
  & = &
 \prod\limits_{s,s\neq r}\left(|z_r-z_s|^{
  \frac{\alpha_s}{\alpha_r}}\right)^{-\frac{\vec p{}^2}{4\alpha}}
  V_{\vec p}^{\rm\scriptscriptstyle T}(z_r,\bar z_r)\ .
\nonumber
\end{eqnarray}
In the limit of large (but finite)
$|\tau_r|$,  each $D_{\tau_r}$ tends to  the small disc $D_{\epsilon_r}$  centered at $z_r$.
The  radius $\epsilon_r$ of $D_{\epsilon_r}$ is given by
$$
{\tau_r\over \alpha}
= \ln \epsilon_r + \sum\limits_{s\neq r}
{\alpha_s \over \alpha_r} \ln |z_r-z_s| \;.
$$
In this limit
the conformal anomaly takes the form
$$
{\cal A}_{D_{\tau_r}} \; = \; \prod\limits_{s,s\neq r}
\left(|z_r-z_s|^{\frac{\alpha_s}{\alpha_r}}\right)^{\frac16},
$$
and one  can rewrite (\ref{amplit_8}) as
\begin{eqnarray}
 \label{amplit_9}
 \Psi_r[X^i_r(\sigma_r)]
&=&
 {\rm e}^{\lambda_r^{\rm\scriptscriptstyle T}\frac{\tau_r}{\alpha_r}}
 \epsilon_r^{-\frac{c^{\rm\scriptscriptstyle T}}{12}}
 \prod\limits_{s,s\neq r}
 \left(|z_r-z_s|^{\frac{\alpha_s}{\alpha_r}}\right)^{\frac{c^{\rm\scriptscriptstyle T}}{12}}
\\
 &&\hspace{30pt}\times
 \int\limits_{D_{\tau_r}}\!
  {\cal D}^{g_{\scriptscriptstyle\rm pl}}
  X^i(z,\bar z)\;
  {\rm e}^{-S[g_{\scriptscriptstyle\rm pl}, X^i(z,\bar z)]}
  \ w_r^*V_r^{\rm\scriptscriptstyle T}(z_r,\bar
  z_r)\ .\nonumber
\end{eqnarray}
The functional integrals over the fields $X^i(z,\bar z)$ on the
regions $M_\tau$ and $D_{\tau_r}\; (r=1,\dots,N),$ integrated over
common boundary values along common boundaries, combine to the
functional integral over the fields on the entire complex plane.
Equations (\ref{amplit_2}), (\ref{anomaly_1}), (\ref{amplit_9})
then imply the following expression:
\begin{eqnarray}
 \label{transverse}
  W^{\scriptscriptstyle\rm T}
  & = & [2\pi\alpha]^{1\over 2}
(  {\cal A}_{M_\tau})^{c^{\rm \scriptscriptstyle T}}
  [\epsilon]^{-\frac{c^{\rm\scriptscriptstyle T}}{12}}
  \prod\limits_{r\neq s}|z_r-z_s|^{\frac{\alpha_s}{\alpha_r}
  \frac{c^{\rm\scriptscriptstyle T}}{12}}\!\\
 &&\hspace{30pt}\times
  \int_{\mathbb{C}}\!\!{\cal D}^{g_{\scriptscriptstyle\rm pl}}
  X^i(z,\bar z)\;{\rm e}^{-S[g_{\scriptscriptstyle\rm pl}, X^i(z,\bar z)]}
  \prod\limits_r\Big(w_r^*V_r^{\rm\scriptscriptstyle T}(z_r,\bar
  z_r)\Big)
  \nonumber \\
  & = &
\delta\left(\sum\limits_{r=1}^N \vec p_r\right)[2\pi\alpha]^{1\over 2}
  ({\cal A}_{M_\tau})^{c^{\rm \scriptscriptstyle T}}
  [\epsilon]^{-\frac{c^{\rm\scriptscriptstyle T}}{12}}
  \prod\limits_{r\neq s}|z_r-z_s|^{\frac{\alpha_s}{\alpha_r}
  \frac{c^{\rm\scriptscriptstyle T}}{12}}\!
  \left\langle
  \prod\limits_r\ w_r^*V_r^{\rm\scriptscriptstyle T}(z_r,\bar z_r)
  \right\rangle^{\rm \scriptscriptstyle T},
    \nonumber
\end{eqnarray}
where  the abbreviated notations (\ref{notation}) are used.

\subsection{Longitudinal sector}

The conformal field theory describing the string excitations in
the Brower sector is not defined by any action
functional\footnote{Note that the Coulomb gas formalism,  although
based on the action functional, does not provide the standard
Lagrangean field-theoretical framework for the minimal models and
should be regarded rather as an ansatz for calculating the
correlation functions.} and Mandelstam's functional method of
constructing amplitudes cannot be applied. There are also no
counterparts of the local fields $\chi^i(z), \widetilde\chi^i(\bar
z)$ and the standard construction of the splitting--joining vertex
based on the connection condition is not directly available. One
can however implement the idea of joining--splitting interaction
defining the light-cone amplitudes  in terms of the conformal
field theory on the light-cone diagram. To this end one can use an
appropriate generalization of formula (\ref{transverse}) derived
in the previous subsection. This leads to the following Brower
sector contribution
\begin{eqnarray}
 \label{Brower}
  W^{\scriptscriptstyle\rm B}
  & = &
({\cal A}_{M_\tau})^{c^{\rm \scriptscriptstyle B}}
  [\epsilon]^{-\frac{c^{\rm\scriptscriptstyle B}}{12}}
  \prod\limits_{r\neq s}|z_r-z_s|^{\frac{\alpha_s}{\alpha_r}
  \frac{c^{\rm\scriptscriptstyle B}}{12}}
  \left\langle
  \prod\limits_r w_r^*V_r^{\rm\scriptscriptstyle B}(z_r,\bar z_r)
  \right\rangle^{\!\!\!\scriptscriptstyle\rm B} \ ,
\end{eqnarray}
where the vertex operators $V_r^{\rm\scriptscriptstyle B}(z_r,\bar
z_r)$ correspond to the external states
$|\,r\,\rangle^{\rm\scriptscriptstyle B}$ in the Brower sector,
and the correlation function is calculated within an appropriate
minimal model. Although we have expressed the Brower contribution
in terms of the CFT on the complex plane, one can easily transform
the theory  back to the light-cone diagram $\Sigma$. In fact, such
transformation can be seen as a rigorous definition of the minimal
model on $\Sigma$.

Let us now turn to the Liouville longitudinal sector. According to
the Liouville interpretation advocated in Subsection 2.3 the
energy--momentum tensor is the only local field in this sector and
the standard state--operator correspondence is replaced by the
state--puncture correspondence. These conclusions were derived by
analyzing the free theory on the cylinder. In order to implement
the idea of splitting--joining interaction  some extension of the
theory from a cylinder to an arbitrary light-cone diagram is
required. As we have seen in the other sectors this can be done in
terms of the Mandelstam map and a well defined conformal theory on
the complex plane. Our strategy is to find all the
properties of such theory which are indispensable for a
construction of the Lorentz covariant string amplitudes. As a
guiding principle one can use the properties of the free theory
already derived in Subsection 2.3 along with the above derivation
of the transverse and the Brower sector contributions to the
string amplitude. This leads to the following assumptions
concerning the Liouville sector.

Our first requirement concerns the conformal anomaly.
We assume that it has its universal form given by the Liouville action  and
depends on the central charge $c^{\rm \scriptscriptstyle L}=1+48\beta$
in the standard way.

The second assumption is related to the fact that there is no
operator--state correspondence in this sector. To the ground state
applied to the free end of a semi-infinite cylinder there
corresponds a
 puncture on the complex plane.
In  the case of tree light-cone diagrams with $N$ external states,
instead of correlation functions of $N$ vertex operators and
arbitrary number of the energy--momentum tensor insertions on the
Riemann sphere $S^2$, one should rather expect correlation
functions of arbitrary number of the energy--momentum tensor
insertions on the Riemann sphere $S^2(z_1,\dots,z_N)$ with $N$
punctures :
\begin{equation}
\label{badnotation}
\left\langle\, \prod_j T^{\rm \scriptscriptstyle L}(w_j)
\prod_k \widetilde T^{\rm \scriptscriptstyle L}(\bar w_k)
\,\right\rangle_{S^2(z_1,\dots,z_N)} \ .
\end{equation}
Since the vacuum of this sector is not $PSL(2,\mathbb{C})$
invariant both the partition function (no energy--momentum tensor
insertions) and the correlations functions of the energy--momentum
tensor on $S^2(z_1,\dots,z_N)$ depend on the locations $z_r$ of
the punctures. For this reason and also because the properties of
the punctures are similar to that of the primary vertex operators
in the standard CFT we shall replace the clumsy notation
(\ref{badnotation}) by
\begin{equation}
\label{Lcorrelators}
\left\langle \,\prod_j T^{\rm \scriptscriptstyle L}(w_j)
\prod_k \widetilde T^{\rm \scriptscriptstyle L}(\bar w_k)
\prod\limits_{r=1}^N P(z_r,\bar z_r)\,\right\rangle^{\!\!\!\rm \scriptscriptstyle L} \ .
\end{equation}
The theory on the cylinder discussed in Subsection 2.3.
defines the Liouville theory on the Riemann sphere with two punctures.
In this case one can easily derive (just from the definitions of the
objects involved) the operator-operator product expansion (OOPE) and
the operator-puncture product expansion (OPPE):
\begin{eqnarray}
\label{TTOPE}
T^{\rm \scriptscriptstyle L}(w)T^{\rm \scriptscriptstyle L}(z)
&=&{\frac12 (1+48\beta)\over (w-z)^4 }
+ {2 \over (w-z)^2}T^{\rm \scriptscriptstyle L}(z)
+{ 1\over w-z}\partial T^{\rm \scriptscriptstyle L}(z) + \dots \ , \\
\label{TPOPE}
T^{\rm \scriptscriptstyle L}(w) P(z, \bar z) &=&
{2\beta \over (w-z)^2}P(z, \bar z)
+{ 1 \over w-z}\partial P(z, \bar z) + \dots \ .
\end{eqnarray}
The new assumption which can be seen as the connection condition
for the energy--momentum tensor is that these expansions also hold
in the case of the sphere with $N$ punctures and that the
punctures are the only singularities of the energy--momentum
tensor.

The third requirement one has to impose on the correlation functions
(\ref{Lcorrelators}) concerns their transformation properties
with respect to the conformal change of their arguments:
\begin{eqnarray}
\label{LTtrans}
\delta_{\epsilon}T^{\rm\scriptscriptstyle L}(z)
&=&- {1\over 2\pi i}\oint dw\, \epsilon(w) T^{\rm\scriptscriptstyle L}(w)
T^{\rm\scriptscriptstyle L}(z)   \ ,\\
\label{LPtrans}
\delta_{\epsilon,\tilde\epsilon}P(z,\bar z) &=&
-{1\over 2\pi i}\oint \,d w\,
 \epsilon(w) T^{\scriptscriptstyle L}(w)P(z,\bar z) \\
 \nonumber
&& - {1\over 2\pi i}\oint \,d \bar w\, \tilde\epsilon(\bar w)
\widetilde T^{\scriptscriptstyle L}(\bar w)P(z,\bar z) \ .
\end{eqnarray}
 With the expansions (\ref{TTOPE}) and (\ref{TPOPE}) one can cast this requirement in the form
of the conformal Ward identities (CWI):
\begin{eqnarray}
\label{Tward}
\left\langle \,T^{\rm \scriptscriptstyle L}(w)
\prod_r P(z_r,\bar z_r)\,\right\rangle^{\!\!\!\rm \scriptscriptstyle L}
&=&\sum_r
\left( {2\beta\over (w-z_r)^2} +{1\over w-z_r}{\partial\over \partial z_r}\right)
\left\langle \,
\prod_r P(z_r,\bar z_r)\,\right\rangle^{\!\!\!\rm \scriptscriptstyle L} \ ,\\
\label{TTward}
\left\langle \,T^{\rm\scriptscriptstyle L}(u)T^{\rm \scriptscriptstyle L}(w)
\prod_r P(z_r,\bar z_r)\,\right\rangle^{\!\!\!\rm \scriptscriptstyle L}
&=&
{\frac12 (1+48\beta)\over (u-w)^4 }\left\langle \,
\prod_r P(z_r,\bar z_r)\,\right\rangle^{\!\!\!\rm \scriptscriptstyle L} \\
\nonumber
&+& \left({2 \over (u-w)^2} + {1\over u-w}{\partial\over\partial w}\right)
\left\langle \,T^{\rm \scriptscriptstyle L}(w)
\prod_r P(z_r,\bar z_r)\,\right\rangle^{\!\!\!\rm \scriptscriptstyle L} \\
\nonumber
&+&\sum_r
\left( {2\beta\over (u-z_r)^2} +{1\over u-z_r}{\partial\over \partial z_r}\right)
\left\langle \,T^{\rm \scriptscriptstyle L}(w)
\prod_r P(z_r,\bar z_r)\,\right\rangle^{\!\!\!\rm \scriptscriptstyle L}
\ .
\end{eqnarray}
Just as in the standard CFT the equations above completely
determine all the three puncture correlation functions
(\ref{Lcorrelators}). In particular in the case of no
energy--momentum insertions (the three puncture partition
function) we have
\begin{equation}
\label{3_puncture}
 \left\langle \, P(z_1,\bar z_1)P(z_2,\bar
z_2)P(z_3,\bar z_3)\,\right\rangle = {C\over |z_1
-z_2|^{2\beta}|z_1 -z_3|^{2\beta}|z_2 -z_3|^{2\beta} }\ .
\end{equation}
Let us stress that the correlation functions of punctures does not
have the interpretation of the vacuum expectation value of some
operators acting in the Hilbert space ${\cal V}^{\rm
\scriptscriptstyle L}$ of the free theory. This means in
particular that we are not forced to require any product expansion
for punctures. Still one can follow most of the constructions of
the standard CFT. In particular the familiar state--operator
correspondence can be replaced by the state--puncture
correspondence defined by
\begin{eqnarray}
\label{statepuncture}
L^{\rm \scriptscriptstyle L}_{-n_1}\dots
L^{\rm \scriptscriptstyle L}_{-n_N} |\,0\,\rangle^{\rm
\scriptscriptstyle L}
 & \longrightarrow  &
 L^{\rm \scriptscriptstyle L}_{-n_1}\dots
 L^{\rm \scriptscriptstyle L}_{-n_N}
 \cdot
 P(z,\bar z)
 \\
 & \equiv &
 {1\over (2\pi i)^N} \oint_{C_1} dz_1
 {T^{\rm \scriptscriptstyle L}(z_1)\over (z_1-z)^{n_1-1}}
 \dots
 \oint_{C_N} dz_N{T^{\rm \scriptscriptstyle L}(z_N)\over (z_N-z)^{n_N-1}}
 P(z,\bar z)\nonumber
 \end{eqnarray}
where the contours  of integration are chosen such that $C_i$ surrounds
$C_{i+1}$ for $i=1,\dots,N-1$, and $C_{N}$ surrounds  the point $z$.
Using this prescription one can associate to each state
$|\,r\,\rangle^{\rm \scriptscriptstyle L}\in {\cal V}^{\rm \scriptscriptstyle L}$
a uniquely determined object
$V^{\rm \scriptscriptstyle L}_r(z,\bar z)$ which we shall call the vertex puncture
corresponding to $|\,r\,\rangle^{\rm \scriptscriptstyle L}$.
Using such  vertex punctures
one can define the Liouville sector contribution to the tree string amplitude
by the formula analogous to those of the Brower (\ref{Brower})
and of the transverse (\ref{transverse}) sectors
\begin{eqnarray}
 \label{Liouville}
  W^{\scriptscriptstyle\rm L}
  & = &
({\cal A}_{M_\tau})^{c^{\rm \scriptscriptstyle L}}
  [\epsilon]^{-\frac{c^{\rm\scriptscriptstyle L}}{12}}
  \prod\limits_{r \neq s}|z_r-z_s|^{\frac{\alpha_s}{\alpha_r}
  \frac{c^{\rm\scriptscriptstyle L}}{12}}
  \left\langle
  \prod\limits_r\ w_r^*V_r^{\rm\scriptscriptstyle L}(z_r,\bar z_r)
  \right\rangle^{\!\!\!\rm \scriptscriptstyle L} \ .
\end{eqnarray}
The correlation function above can be in principle
rewritten in terms of the contour integrals of the correlators (\ref{Lcorrelators}).

\subsection{Light-cone amplitudes}

Gathering formulae (\ref{amplitude}), (\ref{transverse}), (\ref{Brower}), (\ref{Liouville}),
 (\ref{anomaly}) of the previous subsections one gets the following expression
for the light-cone amplitude corresponding to a given type of
light-cone diagram
\begin{eqnarray*}
A_\Sigma &=&\prod\limits_{\mu = 0}^{d-1} \delta\left(\sum\limits_{r=1}^N p_r^\mu \right)
\int\prod\limits_{I=2}^{N-2}d^2(\rho_I-\rho_1)\;[2\pi\alpha]^{1\over 2}
   A^6
 [\alpha]^{-3} [P]^{-2}
 \\
 &\times&
 \prod\limits_{r\neq s}|z_r-z_s|^{2\frac{\alpha_s}{\alpha_r}}
  \left\langle
  \prod\limits_r w_r^*V_r^{\rm\scriptscriptstyle T}(z_r,\bar z_r)
  \right\rangle^{\!\!\!\rm\scriptscriptstyle T}
   \left\langle
  \prod\limits_r w_r^*V_r^{\rm\scriptscriptstyle B}(z_r,\bar z_r)
  \right\rangle^{\!\!\!\rm\scriptscriptstyle B}
   \left\langle
  \prod\limits_r w_r^*V_r^{\rm\scriptscriptstyle L}(z_r,\bar z_r)
  \right\rangle^{\!\!\!\rm\scriptscriptstyle L} \ .
    \nonumber
\end{eqnarray*}
A more symmetric form can be obtained by proceeding to the
Koba--Nielsen variables. We choose  $z_2,\dots,z_{N-2}$ as a new
integration variables. The Jacobian of this change takes the form
(Appendix A.5)
$$
{\cal J} =
\left|(z_N-z_{N-1})(z_N - z_1)(z_{N-1}-z_1)\right|^2  \
A^{-6}{[\alpha]}^2{[P]}^2\ ,
$$
and
\begin{eqnarray}
 \label{lcamplitudesigma}
A_\Sigma &=& (2\pi)^{D\over 2} [\alpha]^{-{1\over 2}}
\prod\limits_{\mu = 0}^{d-1} \delta\left(\sum\limits_{r=1}^N p_r^\mu \right)\\
\nonumber
&\times &
\int\limits_{D_\Sigma} \prod\limits_{r=2}^{N-2}d^2 z_r\;
\left|(z_N-z_{N-1})(z_N - z_1)(z_{N-1}-z_1)\right|^2  \\
 &\times&
 \prod\limits_{r\neq s}|z_r-z_s|^{2\frac{\alpha_s}{\alpha_r}}
  \left\langle
  \prod\limits_r w_r^*V_r^{\rm\scriptscriptstyle T}(z_r,\bar z_r)
  \right\rangle^{\!\!\!\rm\scriptscriptstyle T}
   \left\langle
  \prod\limits_r w_r^*V_r^{\rm\scriptscriptstyle B}(z_r,\bar z_r)
  \right\rangle^{\!\!\!\rm\scriptscriptstyle B}
   \left\langle
  \prod\limits_r w_r^*V_r^{\rm\scriptscriptstyle L}(z_r,\bar z_r)
  \right\rangle^{\!\!\!\rm\scriptscriptstyle L} \ .
    \nonumber
\end{eqnarray}
Let us note that $[\alpha]^{-{1\over 2}} = \prod\limits_{r=1}^n \alpha_r^{-{1\over2}}$
is the expected relativistic flux factor. The integration domain $D_\Sigma$ for
the Koba-Nielsen variables depends on the geometry of the light-cone diagram $\Sigma$.
It is  well known from the theory of the critical string that the domains of integrations
for different types of the tree light-cone diagrams with $N$ external legs sum up
to the whole range of integration $\mathbb{C}^{N-3}$ of the Koba-Nielsen variables.
Hence the full $N$-string tree amplitude is given by formula (\ref{lcamplitudesigma})
with the domain $D_\Sigma$ replaced by $\mathbb{C}^{N-3}$.

In general the calculation of the amplitude
(\ref{lcamplitudesigma}) is difficult due to the complicated
dependence of the transformed vertex operators and vertex
punctures on the Mandelstam map. An essential simplification
occurs  for the (tachionic) ground states. In this case the
transformed vertex operators of the transverse sector satisfy
formula (\ref{w*vertex}). The corresponding formulae in the Brower
and the Liouville sectors read
\begin{eqnarray*}
w_r^*V_{p,q}^{\rm\scriptscriptstyle B}(z_r,\bar z_r)
 & = &
 \prod\limits_{s,s\neq r}|z_r-z_s|^{-
  \frac{\alpha_s}{\alpha_r}2h_m(p,q)}
  V_{p,q}^{\rm\scriptscriptstyle B}(z_r,\bar z_r) \ ,\\
  w_r^*P(z_r,\bar z_r)
 & = &
 \prod\limits_{s,s\neq r}|z_r-z_s|^{ -\frac{\alpha_s}{\alpha_r}4\beta}
 P(z_r,\bar z_r)\ .
\end{eqnarray*}
Taking this into account and using the relation
$$
p^-_r=  {2\sqrt{\alpha}\over \alpha_r}
\left({{\vec p}^{\;2} \over 4 \alpha} + 2h_m(p,q) + 2\beta -2\right)
$$
valid for the ground states one gets
\begin{eqnarray*}
 \prod\limits_{r\neq s}|z_r-z_s|^{\frac{\alpha_s}{\alpha_r}2}
  \left\langle
  \prod\limits_r w^*V_{\vec p_r}^{\rm\scriptscriptstyle T}(z_r,\bar z_r)
  \right\rangle^{\!\!\!\rm\scriptscriptstyle T}
   \left\langle
  \prod\limits_r w^*V_{p,q}^{\rm\scriptscriptstyle B}(z_r,\bar z_r)
  \right\rangle^{\!\!\!\rm\scriptscriptstyle B}
   \left\langle
  \prod\limits_r w^*P(z_r,\bar z_r)
  \right\rangle^{\!\!\!\rm\scriptscriptstyle L} &&\\
 = \;   \prod\limits_{r\neq s}|z_r-z_s|^{-{p^+_sp^-_r\over 2\alpha}}
  \left\langle
  \prod\limits_r V_{\vec p_r}^{\rm\scriptscriptstyle T}(z_r,\bar z_r)
  \right\rangle^{\!\!\!\rm\scriptscriptstyle T}
   \left\langle
  \prod\limits_r V_{p,q}^{\rm\scriptscriptstyle B}(z_r,\bar z_r)
  \right\rangle^{\!\!\!\rm\scriptscriptstyle B}
   \left\langle
  \prod\limits_r P(z_r,\bar z_r)
  \right\rangle^{\!\!\!\rm\scriptscriptstyle L} &&\ .
\end{eqnarray*}
Calculating the correlator in the transverse sector one gets an
explicitly covariant expression for the tachionic tree string
amplitude:
\begin{eqnarray}
 \label{tachions}
A &=& (2\pi)^{D\over 2} [\alpha]^{-{1\over 2}}
\prod\limits_{\mu = 0}^{d-1} \delta\left(\sum\limits_{r=1}^N p_r^\mu \right)\\
\nonumber
&\times &
\int\limits_{\mathbb{C}^{N-3}} \prod\limits_{r=2}^{N-2}d^2 z_r\;
\left|(z_N-z_{N-1})(z_N - z_1)(z_{N-1}-z_1)\right|^2  \\
 &\times&
 \prod\limits_{r\neq s}|z_r-z_s|^{{p_s\cdot p_r\over 2\alpha}}
 \left\langle
  \prod\limits_r V_{p,q}^{\rm\scriptscriptstyle B}(z_r,\bar z_r)
  \right\rangle^{\!\!\!\rm\scriptscriptstyle B}
   \left\langle
  \prod\limits_r P(z_r,\bar z_r)
  \right\rangle^{\!\!\!\rm\scriptscriptstyle L}\ .
    \nonumber
\end{eqnarray}

\section{Lorentz covariance}

\setcounter{equation}{0}

In this section we shall analyze Lorentz covariance of the tree
string amplitudes for arbitrary states. Our approach is a slight
modification of the method originally invented by Mandelstam
\cite{Mandelstam74} and further developed in
\cite{hikko,Sin88,kisa}. Nontrivial part concerns the generators
$M^{i-}$ and we restrict ourselves to this case.

Our aim is to calculate the effect of an infinitesimal Lorentz
rotation
\begin{equation}
\label{Lrot} |r\,\rangle \rightarrow |r\,\rangle  + \varepsilon
M^{i-}|r\,\rangle
\end{equation}
on the tree string amplitude. To this end we shall derive a
convenient  expression for the Lorentz generators $M^{i-}$
\cite{hikko,Sin88,kisa}. Applying the time splitting technique and
the $\zeta$ function regularization one gets
\begin{eqnarray}
\label{Gamma} \Gamma^i(\tau)
 &  \equiv &
\lim\limits_{\epsilon \to 0} \oint\limits_\tau \frac{d\rho}{\pi
i}\; {\cal T}\left(T(\rho +\epsilon)\chi^i(\rho)\right)\\
\nonumber
 &  = & \frac12\frac{\sqrt{\alpha}}{p^+}\left[\left(L_0 - \frac{c}{24}\right)q_0^i +
q_0^i \left(L_0 - \frac{c}{24}\right) \right]
 \\ \nonumber
&& -\; i\frac{\sqrt\alpha}{p^+}\sum_{n=1}^\infty \frac{1}{n}
\left( a_{-n}^i L_n - L_{-n} a_n^i\right) -i\frac{P^i}{p^+}
\oint\limits_\tau \frac{d\rho}{2\pi i}\; \rho\,T(\rho)\ ,\nonumber
\end{eqnarray}
where $ T(\rho)\equiv T^{\rm\scriptscriptstyle T}(\rho)
+T^{\rm\scriptscriptstyle B}(\rho)+T^{\rm\scriptscriptstyle
L}(\rho) $ is the total energy--momentum tensor,
 $\cal T$ means (Euclidean) time ordering, and the integration goes
over the contour of constant time $\tau$. Then for every external
leg one can write $M^{i-}_r$ in the form
\begin{equation}
 \label{M_iminus1} M^{i-}_r \; = \;\Gamma^i_r(\tau)_r +
\widetilde\Gamma^i_r(\tau) + {iP^i_r\over
\sqrt\alpha}{\!\partial\over \partial \alpha_r} + {iP^i_r\over
\sqrt\alpha} \left( Q_r(\tau) + \widetilde Q_r(\tau) \right)\ ,
\end{equation}
 where
\begin{eqnarray}
\label{Qrho} Q_r(\tau)
 & =&
\oint\limits_\tau \frac{d\rho}{2\pi i } \;{\rho\over \alpha_r}
T(\rho) \ ,
\end{eqnarray}
and $\widetilde \Gamma^i_r(\tau), \widetilde Q_r(\tau)$ are the
 antiholomorphic counterparts of
$\Gamma^i_r(\tau), Q_r(\tau)$, respectively. Let us stress that
although $M^{i-}$ is conserved the decomposition (\ref{M_iminus1})
depends on the choice of the contour. One can however verify that
the decomposition (\ref{M_iminus1}) is consistent with the free
string evolution, {\em i.e.}
\begin{eqnarray*}
{\rm e}^{-{1\over 2\sqrt{\alpha}}P^-_r(\tau-\tau')}
\Gamma^i_r(\tau') &=&\Gamma^i_r(\tau)
{\rm e}^{-{1\over 2\sqrt{\alpha}}P^-_r(\tau-\tau')}\ ,\\
{\rm e}^{-{1\over 2\sqrt{\alpha}}P^-_r(\tau-\tau')} Q_r(\tau')
&=&Q_r(\tau) {\rm e}^{-{1\over 2\sqrt{\alpha}}P^-_r(\tau-\tau')}\
.
\end{eqnarray*}
These formulae hold also in the interacting theory if we assume
the following transformation rule of the local fields $A(\rho,\bar
\rho)$ on the light-cone diagram with respect to the infinitesimal
conformal transformations:
\begin{eqnarray}
\label{transfo} \delta_{\epsilon,\tilde\epsilon}A(\rho,\bar \rho)
&=& -\oint {d \omega \over 2\pi i}
 \epsilon(\omega) T(\omega)A(\rho,\bar \rho)
-\oint {d \bar \omega\over 2\pi i} \tilde\epsilon(\bar \omega)
\widetilde T(\bar \omega)A(\rho,\bar \rho) \ .
\end{eqnarray}
It is certainly satisfied in the sectors described by standard
CFT. As we have discussed in the previous section it is assumed
in the Liouville sector as well. With this assumption the action
of the operators $M_r^{i-}-{i\over \sqrt{\alpha}}  P^i{\partial
\over \partial \alpha_r}$ on each external state $|r\,\rangle$
can be replaced by the insertions of the operators
$\Gamma^i_r(\tau_{r}'),\widetilde\Gamma^i_r(\tau_{r}'),
Q_r(\tau_r''),\widetilde Q_r(\tau_r'')$ at arbitrary chosen times
$\tau_r',\tau_r''$ on the corresponding external legs.

Let us first calculate the effect of
$\Gamma^i_r(\tau_{r}'),\widetilde\Gamma^i_r(\tau_{r}')$
insertions. To the sum of contributions of  external legs one can
add pairs of opposite contributions of opposite oriented contours
for each intermediate cylinder. Since the insertion times are
arbitrary one can move the contours of integration towards nearest
interaction times $\tau_I$ in such a way that each interaction
point $Z_I$ is surrounded by the contour $C_I$ consisting of three
disjoint components $C_{I,s}$ as indicated on Fig.~1.

\vskip 5mm

\includegraphics*[width=.9\textwidth]{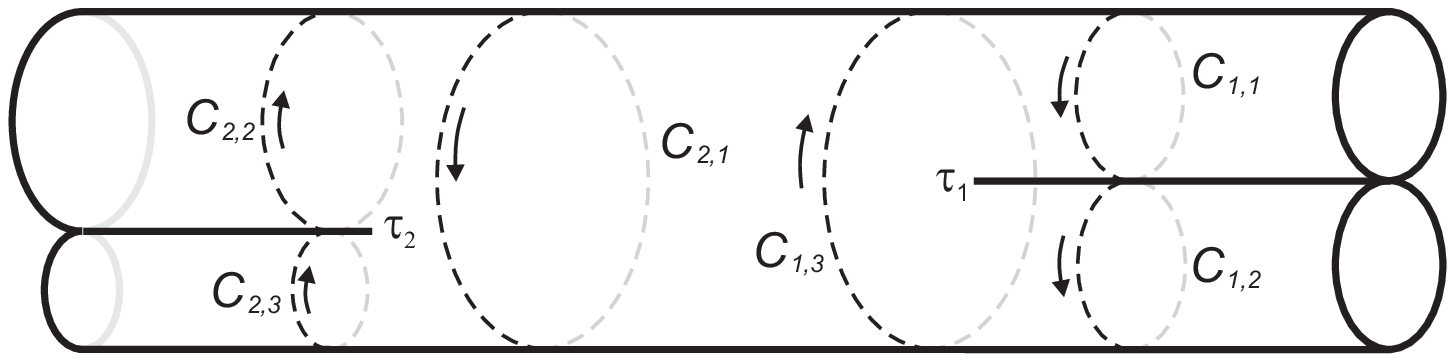}

\vskip 2mm

\centerline{\small{\bf Fig.~1}~ Integration contours on the
light-cone graph}

\vskip 3mm

The contribution of each interaction point can be rearranged as
follows:
\begin{eqnarray}
\nonumber
 \Gamma_I^i + \widetilde\Gamma_I^i &=&
\oint\limits_{C_I} \frac{d\rho}{\pi i} {\cal
T}\left(T(\rho)\chi^i(\rho)\right)+ \oint\limits_{C_I}
\frac{d\bar\rho}{\pi i}
{\cal T}\left(\widetilde T(\bar\rho)\widetilde\chi^i(\bar\rho)\right)\\
\nonumber &=& \Upsilon_I^i + \widetilde\Upsilon_I^i +
\sum\limits_{s=1}^3 {1 \over 2\sqrt{\alpha} }{\cal T}\left(
 P^-_s (\chi^i(\rho_I)+\widetilde\chi^i(\bar\rho_I))\right) \ ,\\
 \label{deco}
\Upsilon_I^i &\equiv& \oint\limits_{C_I} \frac{d\rho}{\pi
i}\,{\cal T}\left(T(\rho) \left(\chi^i(\rho)-
\chi^i(\rho_I)\right)\right) \ .
\end{eqnarray}
In order to calculate the contour integrals
$\Upsilon_I^i,\widetilde\Upsilon_I^i$ one can proceed  from the
light-cone diagram to the complex plane by the Mandelstam map,
\begin{eqnarray*}
\nonumber \Upsilon_I^i & = &
\oint\limits_{\rho^{-1}(C_I)}\frac{dz}{\pi i}\;\frac{1}{\rho'(z)}
{\cal T}\left(T^{\rm \scriptscriptstyle T}(z)\left(\chi^i(z)-\chi^i(z_I)\right)\right)\\
& -& \frac{c^{\rm \scriptscriptstyle T} +c^{\rm
\scriptscriptstyle B} +c^{\rm \scriptscriptstyle L}}{12}
\oint\limits_{\rho^{-1}(C_I)}\frac{dz}{\pi i}\;\frac{1}{\rho'(z)}
\left(
\frac{\rho^{(3)}(z)}{\rho'(z)}-\frac32\frac{\left(\rho''(z)\right)^2}{\left(\rho'(z)\right)^2}
\right)\left(\chi^i(z) - \chi^i(z_I)\right)\ ,
\end{eqnarray*}
and then deform the  integration contour  to one surrounding the
point of interaction $Z_I =\rho^{-1}(\rho_I)$ (Fig.~2).

\vskip 5mm

\includegraphics[width=.9\textwidth]{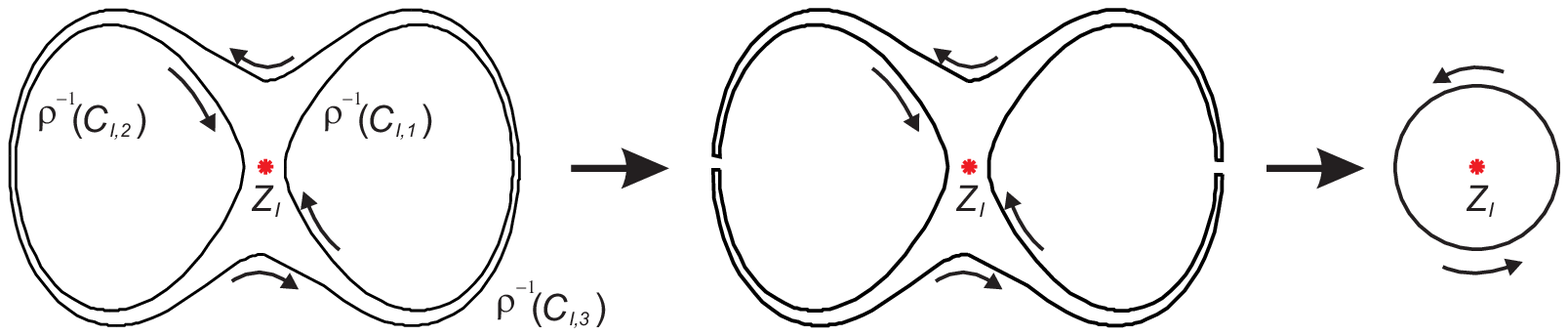}

\vskip 2mm

\centerline{\small{\bf Fig.~2}~ Deformation of the integration
contour on the complex $z$ plane}

\vskip 3mm

As it is shown in Appendix C,
\begin{equation}
\label{Sigma} \Upsilon_I^i = \left(1- \frac{c^{\rm
\scriptscriptstyle T} +c^{\rm \scriptscriptstyle B} +c^{\rm
\scriptscriptstyle L}}{24}\right) \frac{2b}{a^2}\
\partial\chi^i(z_I) + \left(\frac{c^{\rm \scriptscriptstyle
T}+c^{\rm \scriptscriptstyle B} +c^{\rm \scriptscriptstyle
L}}{16}-\frac32\right) \frac{2}{a}\
\partial^2\chi^i(z_I)\ ,
\end{equation}
where the constants $a,b$ are determined by the expansion
$$
\rho(z) - \rho(z_I) = \frac12a(z-z_I)^2 + \frac13b(z-z_I)^3 +
{\cal O}\left((z-z_I)^4\right)
$$
of the Mandelstam map (\ref{Mmap}) near the interaction point.
Since $c^{\rm \scriptscriptstyle T} +c^{\rm \scriptscriptstyle B}
+c^{\rm \scriptscriptstyle L}= 24$ and the calculations in the
antiholomorphic sector are essentially identical, the first two
terms  in (\ref{deco}) vanish. Due to the time ordering involved,
the third term in (\ref{deco}) can be seen as a commutator of the
Hamiltonian
\begin{equation}
\nonumber P^-= \sqrt{\alpha}\oint\limits_{\tau} \frac{d\rho}{\pi
i}\; T(\rho) + \sqrt{\alpha}\oint\limits_{\tau}
\frac{d\bar\rho}{\pi i}\; \widetilde T(\bar\rho)
\end{equation}
with the operator
$$
X^i(\rho_I,\bar\rho_I)={1 \over 2\sqrt{\alpha} }
(\chi^i(\rho_I)+\widetilde\chi^i(\bar\rho_I)) \ .
$$
Thus inside  the correlator  defining the string amplitude one
gets
\begin{eqnarray}
\label{timeder} \sum\limits_{r=1}^N ( \Gamma_r^i(\tau_r) +
\widetilde\Gamma_r^i(\tau_r)) &=& \sum\limits_{I=1}^{N-2}
(\Gamma_I^i + \widetilde\Gamma_I^i) \;=\;
2\sqrt{\alpha}\sum\limits_{I=1}^{N-2}  {\partial\over\partial
\tau_I} X^i(\rho_I,\bar\rho_I)\ .
\end{eqnarray}

Let us now consider the effect of the third  term in the
decomposition (\ref{M_iminus1}). Since the external states are
eigenstates of the momenta operators one can replace $P^i_r$ by
corresponding eigenvalues $p^i_r$. In the $p^+$-sector
 the third term of (\ref{M_iminus1})
is given by the operator ${i\over \sqrt \alpha}\sum_r
p_r^i\frac{\!\partial}{\partial\alpha_r}$ applied to the delta
function $\delta(\sum_r p^+_r)$ and vanishes by the conservation
of $P^i$  momentum. In the other sectors the string amplitude
depends on $\alpha_r$ only via the geometry of the light-cone
diagram. For all $r$ the change $\alpha_r \to \alpha_r
+\delta\alpha_r$ corresponds to the rescaling of the $r^{\rm th}$
external leg
\begin{equation}
\label{rescaling} \rho \to \rho +
\delta\alpha_r{\rho\over\alpha_r} \;\;\;,\;\;\; \bar\rho \to
\bar\rho + \delta\alpha_r{\bar\rho\over\alpha_r}\ .
\end{equation}
It follows from (\ref{transfo}) that the the effect of these
rescaling on the local operators (punctures) representing external
state on the $r^{\rm th}$ leg  is  given by  the insertion of the
operator $- Q_r(\tau)-\widetilde Q_r(\tau)$. Hence inside the
correlator defining the string amplitude for every $r$
$$
{\partial \over \partial \alpha_r} = - Q_r(\tau)-\widetilde
Q_r(\tau)
$$
and therefore the third and the fourth term in (\ref{M_iminus1})
cancel each other.

The net effect of the infinitesimal  Lorentz rotation (\ref{Lrot})
on the contribution $A_\Sigma$ to the tree string amplitude thus
reads
$$
\delta A_\Sigma =2\sqrt{\alpha}\,
\delta\!\left(\sum\limits_{r=1}^N p_r^+\right)
\delta\!\left(\sum\limits_{r=1}^N p_r^-\right)
\int\prod\limits_{I=2}^{N-2}d^2(\rho_I-\rho_1)
\sum\limits_{I=1}^{N-2}{\partial\over \partial \tau_I}
   (W^{\scriptscriptstyle\rm T}[X^i_I]
    W^{\scriptscriptstyle\rm B}W^{\scriptscriptstyle\rm L})\ ,
$$
where $W^{\scriptscriptstyle\rm T}[X^i_I]$ denotes the transverse
sector contribution modified by the insertion of the operator
$X^i(\rho)$ at the interaction point $\rho_I$. Due to the ordering
of the interaction times for a given  light-cone diagram the
integral above yields non-vanishing boundary terms. It is well
known from the theory of critical string \cite{crge,kaki,hikko}
that boundaries of integration of all light-cone diagrams
contributing to a given tree string amplitude can be arranged into
pairs of identical boundaries corresponding to different diagrams.
An example of such a pair is exhibited on Fig.~3.

\vskip 5mm

\includegraphics*[width=.9\textwidth]{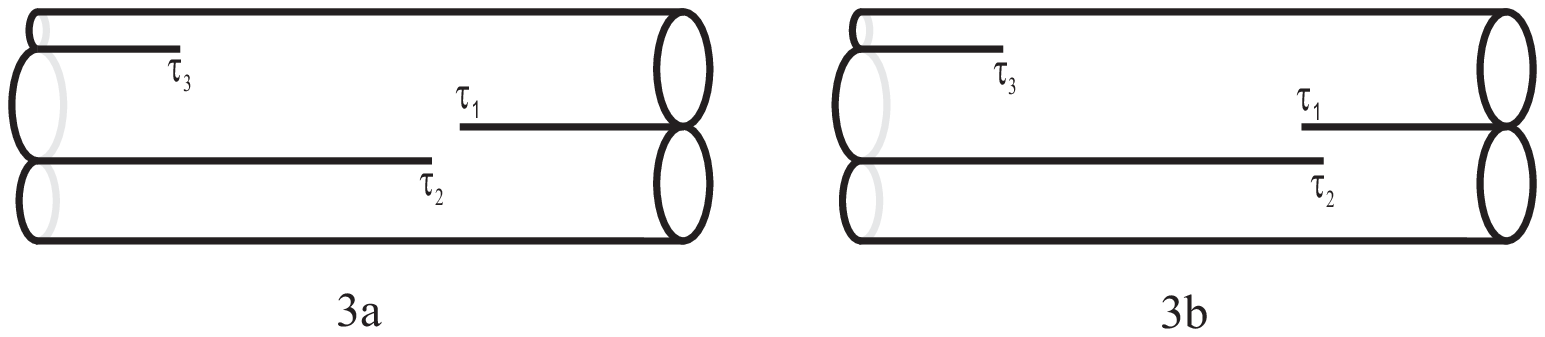}

\vskip 2mm

\centerline{\small{\bf Fig.~3}~ $s$-channel (3a) and $t$-channel
(3b) boundary contribution to the scattering amplitude}

\vskip 3mm

The boundary terms corresponding to each of these pairs cancel
each other if the expressions $W^{\scriptscriptstyle\rm T}[X^i_I]
W^{\scriptscriptstyle\rm B}W^{\scriptscriptstyle\rm L}$ for
different diagrams are the same on the common boundary component.
In our construction, where the expression
$W^{\scriptscriptstyle\rm T}[X^i_I] W^{\scriptscriptstyle\rm
B}W^{\scriptscriptstyle\rm L}$  was defined in terms of a
Mandelstam map and conformal field theories on the complex plane,
this condition is satisfied. Indeed the Mandelstam map is the same
for light-cone diagrams with common boundaries of integration and
the correlators on the complex plane  depend on the interaction
times only through the insertion $X^i(Z_I)$. This completes the
proof of the Lorentz covariance of the light-cone tree string
amplitudes introduced in Sect.~3.

\section{Unitarity}

\setcounter{equation}{0}

Our derivation of the tree light-cone amplitudes presented in Section 3 shows that
the propertirs of the Liouville sector listed in the Introduction:
the spectrum, the conformal anomaly, and the
conformal Ward identities form a minimal set of assumptions
necessary for implementing the idea of splitting--joining interaction
for all massive string models of the bosonic discrete series.
It follows from our calculations in Section 4 that these properties are
also sufficient for the Lorentz covariance of the light-cone amplitudes.
 The only fundamental property of standard CFT we have
not yet required in the Liouville sector is the puncture --
puncture product expansion (PPPE). The question arises to what
extend this property is necessary for the perturbative unitarity
of tree amplitudes.

We shall start with a brief discussion  of the critical string.
According to the presentation of Sect.~3 the tree  amplitudes can
be constructed by means of correlators of $24$ copies of the
scalar conformal field on the complex plane. In order to prove the
unitarity one has to identify these amplitudes as terms of a Dyson
perturbative expansion in the space of multi-string states. Such
identification is easily seen in the functional approach where one
can  express the amplitudes in terms of path integrals on  light
cone diagrams.  Cutting the path integral  open just before and
just after each interaction time one gets the free string
propagation of multi-string states between instant
joining--splitting interaction vertices (Fig.~4). This intuitive
picture can be justify by explicit calculations. Considering the
limit of an infinitely short pants one can construct an operator
corresponding to the elementary joining--splitting vertex
\cite{crge,kaki,hotuco,grsc83}. This vertex and the free string
evolution yield the Dyson expansion required.

\vskip 5mm

\includegraphics*[width=.9\textwidth]{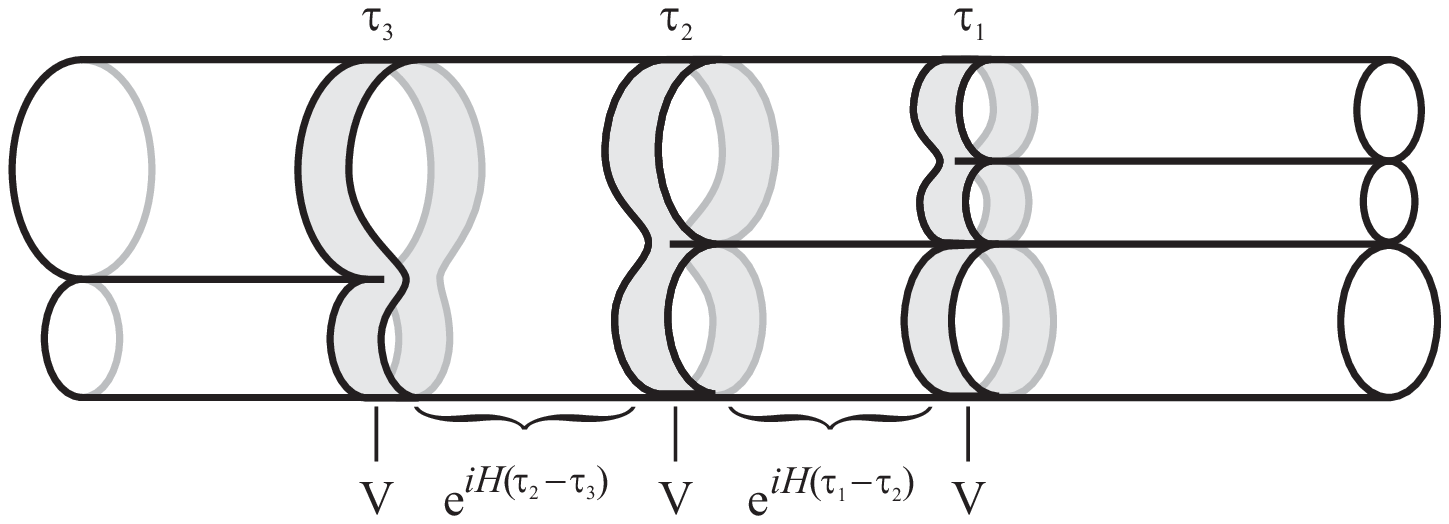}

\vskip 3mm

\centerline{\small{\bf Fig.~4} Decomposition of the light-cone
graph onto vertices and free propagation}

\vskip 2mm

One can extend this line of reasoning  to the string degrees of
freedom  described by arbitrary standard CFT. This concerns in
particular various compactifications of critical string. What we
need to separate the correlator into regions of free  propagation
and interaction vertices, is  a well defined cutting-open
procedure {\em i.e.} a prescription which for any closed curve
with no self-intersections yields  a unique decomposition of the
correlator into scalar product of two states from the free string
Hilbert space. With the operator--state correspondence assumed the
existence of such procedure is equivalent to the operator product
expansion (OPE) which is a fundamental property of any standard
CFT. The detailed  construction of the decomposition of the
correlator into cylinders of free propagation and infinitely short
pants of joining--splitting vertices goes beyond the scope of the
present paper. Let us only mention that one can use for instance
the techniques of \cite{Sonoda:1988mf} where a similar
decomposition into 3-point functions was derived.

Let us now turn to the non-critical string. In the transverse and
in the Brower longitudinal sector one can decompose the
correlators into interaction vertices and free evolution just as
in the case of critical string. A similar decomposition in the
Liouville sector requires the puncture-puncture product expansion
(PPPE). This however leads  to some consistency conditions. One
can  for instance apply PPPE to calculate 4-puncture correlator in
three different ways. The condition that the results should
coincide is usually referred to as the crossing symmetry or the
bootstrap equation. One of its consequences in standard CFT is the
restriction on  possible central charge, conformal dimensions, and
fusion rules known as Vafa's condition \cite{Vafa88}. Using
Lewellen's  derivation of this condition \cite{Lewellen89} one can
show   that the PPPE assumption is in contradiction with the
spectrum of conformal weights in the Liouville sector \cite{hadjas}.

Let us first assume that PPPE holds. Then the spectrum in the
Liouville sector must be essentially larger. In  commonly adopted
approaches to the quantum Liouville theory it is actually
continuous. If we extend the space of external
states to accommodate such spectrum the unitarity will be preserved
but the original physical content of the theory will be lost.
Indeed the extension would result in the continuous family of intercepts.
The only way out is the truncation of  spectrum to a discrete
subset as was proposed for instance in \cite{Gervais:1990be}.
On the other hand one could preserve the space of external states
determined by the free string models and use the Liouville theory with
the continuous spectrum to calculate string amplitudes. In this case
the proof of unitarity known from the critical string theory breaks down
and one has to analyze the problem on the level of string amplitudes
rather than on the level of 2-dim field theories on the light-cone
diagrams. It is an interesting open problem whether the ODZZ proposal
to solve the Liouville theory \cite{Dorn:1994xn,Zamolodchikov:1995aa}  leads to the non-critical string amplitudes
satisfying the conditions of perturbative unitarity.

The second possibility is to relax the PPPE requirement.
Up to our knowledge the only  approach to the Liouville theory
where PPPE is not explicitly or tacitly assumed is the geometric formulation of the 2-dim
quantum gravity \cite{Polyakov82,Takhtajan:1993vt,Takhtajan:zi,Takhtajan:1994vt,Takhtajan:1995fd}.
The Liouville correlation functions
are defined in this approach in terms of path integral over conformal class of Riemannian metrics
with prescribed  singularities at the punctures.
The case of the metrics with parabolic singularities was
extensively analyzed by means of the perturbation expansion around
the classical hyperbolic geometry
\cite{Takhtajan:1993vt,Takhtajan:zi,Takhtajan:1994vt}. In
particular the conformal weight of puncture and the central charge
were calculated, and the conformal Ward identities were proved.
The results are in perfect agreement with the properties of the
Liouville sector we required in Subsections 2.3 and 3.3 on
different grounds. Let us   stress that the  geometrical approach
is strongly justified by the fact that many of its geometric
predictions  can be rigorously proved
(\cite{Takhtajan:zi,Takhtajan:1994vt}, and references therein).
Still it is very far from its final solution. The most
important open questions are the factorization, and the relation
to the ODZZ approach.

Our discussion shows that the relation between the properties of the 2-dim
theory on the world sheet and the perturbative unitarity of the string amplitudes
is in the Liouville longitudinal sector much more complicated than in other
sectors described by standard CFT. It seems that further investigations of this relation should
provide a new insight into the longstanding problem of string interactions in 4 dimensions.

\setcounter{equation}{0}
\renewcommand{\thesection}{A.}
\renewcommand{\thesubsection}{A.\arabic{subsection}}
\renewcommand{\theequation}{A.\arabic{equation}}
\section{The Mandelstam map}
\subsection{Conformal factor of the  Mandelstam map}
Conformal factor of the full Mandelstam map
$$
\rho(z) =\sum\limits_{r=1}^{N}\alpha_r \ln (z-z_r)\ ,
\hskip 5mm
\sum\limits_{r=1}^N \alpha_r= 0\ ,
$$
is
\begin{equation}
\label{full_1}
\sigma = \ln \left|{\partial \rho \over \partial z}\right|^2 = \ln
\left|\sum\limits_{r=1}^{N} {\alpha_r \over z- z_r}\right|^2.
\end{equation}
The metric $\rho^*g_c$ is flat except at singular points
$\{z_r\}_{r=1}^N$ and $\{Z_I\}_{I=1}^{N-2}$. Singularities at
these points  can be calculated by means of the Gauss-Bonnet
theorem
$$
{1\over 4\pi }\int\limits_M\!\sqrt g\,d^2z\, R_g +
{1\over 2\pi }\int\limits_{\partial M}\!ds\, \kappa_g = 2-2g -b \ .
$$
One gets
$$
\sqrt{\rho^*g_c}\ R_{\rho^*g_c}
\;  = \;
-4\pi\sum_{I=1}^{N-2} \delta(z-Z_I)
+4\pi\sum_{r=1}^{N} \delta(z-z_r)\ .
$$
Since
$$
\sqrt{\rho^*g}\ R_{\rho^*g}
\; = \;
-\partial_a \sqrt{g}g^{ab}\partial_b \phi + R_g\ ,
$$
the conformal factor satisfies
$$
-\nabla^2 \sigma
\; =\;
-4\pi\sum_{I=1}^{N-2} \delta(z-Z_I)
+4\pi\sum_{r=1}^{N} \delta(z-z_r)\ ,
\hskip 5mm
\nabla^2\equiv \partial_a\partial^a
= 4\partial_z\partial_{\bar z}\ .
$$
On the complex plane an identity
$$
\nabla^2 \ln |z-w|^2
\; = \;
4\pi \delta(z-w)
$$
holds and the general solution to the previous equation can be written as
\begin{equation}
\label{full_2}
\sigma
\; = \;
\sigma_0 +\sum_{I=1}^{N-2} \ln|z - Z_I|^2 -\sum_{r=1}^{N} \ln|z-z_r|^2\ ,
\end{equation}
where $\sigma_0$ is an arbitrary solution of the homogeneous equation
$$
\nabla^2\sigma_0
\; = \; 0\ .
$$
One can find $\sigma_0$ analyzing the asymptotic behavior of
$\sigma$ for $|z|^2 \to \infty$. From (\ref{full_1}) one gets
\begin{equation}
\label{full_3}
\sigma \; =\;  - 2 \ln |z|^2 + \ln \left|z^2\sum\limits_{r=1}^N
\left({\alpha_r\over z} + {\alpha_r z_r\over(z-z_r)z} \right)\right|^2
\; = \;
- 2 \ln |z|^2 + \ln \left|\sum\limits_{r=1}^N
\alpha_r z_r\right|^2 + o\left({\textstyle {1\over|z|}}\right)\ ,
\end{equation}
while (\ref{full_2}) implies
\begin{equation}
\label{full_4}
\sigma
\; = \;
\sigma_0 -2 \ln|z|^2 + o({\textstyle {1\over|z|}})\ .
\end{equation}
Comparison of  (\ref{full_3}) and (\ref{full_4}) yields $\sigma_0
= \ln A^2$ where
\begin{equation}
\label{A}
A \; \equiv \; \left|\sum\limits_{r=1}^N\alpha_r z_r\right|\ ,
\end{equation}
and leads to  the following expression for the conformal factor
\begin{equation}
\label{full_5}
\sigma =\ln A^2  +\sum_{I=1}^{N-2} \ln|z - Z_I|^2 -\sum_{r=1}^{N} \ln|z-z_r|^2\ .
\end{equation}

\subsection{Boundary terms}
We shall show that
\begin{eqnarray}
\label{global_r}
{1\over 4\pi} \int\limits_{\partial M_r}\!ds\,\sigma n^a\partial_a \sigma
& = &
\ln A^2
- \ln\epsilon_r^2
- \sum\limits_{s(s\neq r)} \ln |z_r-z_s|^2
+ \sum_{I} \ln |Z_I-z_r|^2
\\
\label{local_r}
& = &
2 \ln|\alpha_r | -2 \ln \epsilon_r \ ,\\
\label{global_I}
{1\over 4\pi}\int\limits_{\partial M_I}\!ds\, \sigma n^a\partial_a \sigma
& = &
-\ln A^2
- \ln\epsilon_I^2
- \sum\limits_{J(J\neq I)} \ln |Z_I-Z_J|^2
 +  \sum_{r} \ln |z_r-Z_I|^2\\
 \label{local_I}
& = &
-2\ln c_I - 2\ln \epsilon_I
\; = \;
-\ln c_I - \ln 2r_I\ ,\\
\label{local_0}
{1 \over 8\pi}
\int\limits_{\partial M_\infty}\!ds\, \sigma n^a\partial_a \sigma
& = &
4\ln\epsilon_\infty - 2\ln A
\; = \;
-4\ln r_0 + 2\ln A\ ,
\end{eqnarray}
where $\partial M_r$ and $\partial M_I$ denote small circles of
the radii $\epsilon_r$ and $\epsilon_I$, surrounding points $z_r$
and $Z_I$, respectively, and $\partial M_\infty$ is a big circle
of the radius $\epsilon_\infty$ ``surrounding'' $\infty$.

\vskip 3mm
\noindent  {\bf External points}
\vskip 3mm

\noindent
In the limit
$$
\epsilon_r \; =\;  |z-z_r| \; \rightarrow\; 0
$$
one has
\begin{eqnarray*}
\rho(z)
& \approx &
\alpha_r\ln (z-z_r)\ ,
\\
\sigma
& = &
2 \ln\left|{\partial \rho\over \partial z_r} \right|
\; \approx \;
2 \ln\left| {\alpha_r \over z - z_r}\right|
\; =\;
2 \ln{|\alpha_r |\over \epsilon_r}\ ,
\\
n^a\partial_a\sigma
& = &
 -{\partial \over \epsilon_r}\left(2 \ln{|\alpha_r |\over \epsilon_r}\right)
\; = \;
{2\over\epsilon_r}\ ,
\end{eqnarray*}
and
\begin{eqnarray*}
{1\over 4\pi} \int\limits_{\partial M_r}\!ds\,\sigma n^a\partial_a \sigma
& \approx &
2 \ln|\alpha_r | -2 \ln \epsilon_r\ ,
\end{eqnarray*}
what proves (\ref{local_r}). To demonstrate validity of (\ref{global_r})
we use  for $\sigma$ the expression (\ref{full_5}) with the result:
\begin{eqnarray*}
{1\over 4\pi}\int\limits_{\partial M_r} ds \sigma n^a\partial_a \sigma
&\approx &
{1\over 4\pi}\int\limits_{0}^{2\pi} \epsilon_r d\varphi \ln A^2{2\over \epsilon_r}
- {1\over 4\pi}\int\limits_{0}^{2\pi} \epsilon_r d\varphi \ln \epsilon_r^2{2\over \epsilon_r}\\
&+&{1\over 4\pi}\int\limits_{0}^{2\pi} \epsilon_r d\varphi
\left(-\sum\limits_{s(s\neq r)} \ln |z - z_s|^2 +
 \sum\limits_{I} \ln |z - Z_I|^2\right)
 {2\over \epsilon_r}\\
&\approx&
\ln A^2-  \ln\epsilon_r^2
- \sum\limits_{s(s\neq r)} \ln |z_r-z_s|^2
 +  \sum_{I} \ln |z_r-Z_I|^2\ .
\end{eqnarray*}

\vskip 3mm
\noindent
{\bf  Interaction points}
\vskip 3mm

\noindent
For
$$
\epsilon_I \; = \;  |z-Z_I| \; \rightarrow \; 0
$$
one has
\begin{eqnarray*}
\rho(z)-\rho(Z_I)
& \approx &
{1\over 2}
{\partial^2 \rho\over \partial z^2 }(Z_I)(z-Z_I)^2 ,
\\
{\partial \rho \over \partial z}(z)
& \approx &
{\partial^2 \rho\over \partial z^2 }(Z_I)(z-Z_I)\ .
\end{eqnarray*}
Introducing
$$
c_I \; = \; \left|{\partial^2 \rho\over \partial z^2 }(Z_I)
\right|\ ,
\hskip 5mm
r_I \; = \; |\rho(z)-\rho(Z_I)|\ ,
$$
one gets
$$
r_I \; = \; {1\over 2} c_I \epsilon_I^2
\hskip 3mm \Rightarrow \hskip 3mm
\epsilon_I^2 \; = \; {2 r_I\over c_I}\ ,
$$
and
\begin{eqnarray*}
\sigma
& = &
2 \ln\left|{\partial \rho\over \partial z} \right|
\; \approx \;
2\ln  c_I + 2\ln \epsilon_I
\; = \;
2\ln c_I + \ln {2r_I\over c_I}
\; = \;
\ln c_I +\ln 2r_I \ ,
\\
n^a\partial_a\sigma
& = &
 -{\partial \over \partial\epsilon_r}
(2 \ln c_I + 2\ln \epsilon_I)
\; = \;
- {2\over \epsilon_I}\ .
\end{eqnarray*}
Thus
$$
{1\over 4\pi}\int\limits_{\partial M_I} ds\, \sigma n^a\partial_a \sigma
\; \approx\;
 -2\ln c_I - 2\ln \epsilon_I
\; = \;
-\ln c_I - \ln 2r_I\ ,
$$
what proves (\ref{local_I}).
On the other hand, with the help of (\ref{full_5}) we have
\begin{eqnarray*}
\int\limits_{\partial M_I}\!ds\, \sigma n^a\partial_a \sigma
& \approx &
- \int\limits_{0}^{2\pi} \epsilon_I d\varphi\, \ln A^2 {2\over \epsilon_I}
- \int\limits_{0}^{2\pi} \epsilon_I d\varphi\, \ln \epsilon_I^2{2\over \epsilon_I}
\\
& - &
\int\limits_{0}^{2\pi} \epsilon_I d\varphi
\left(\sum\limits_{J(J\neq I)} \ln |z - Z_I|^2 - \sum\limits_{r} \ln |z - z_r|^2\right)
 {2\over \epsilon_I}
\\
& \approx &
 - 4\pi\ln A^2- 4\pi \ln\epsilon_I^2
-4\pi \sum\limits_{J(J\neq I)} \ln |Z_I-Z_J|^2
 + 4\pi \sum_{I} \ln |Z_I-z_r|^2\ .
\end{eqnarray*}

\noindent {\bf  Point at infinity}
\vskip 3mm

\noindent In the limit
$$
\epsilon_\infty = |z| \to \infty
$$
one has
\begin{eqnarray*}
\rho(z) &=& \sum_{r=1}^N \alpha_r \ln (z-z_r)
\; = \;
\sum_{r=1}^N \alpha_r \ln (1-{z_r\over z})
\\
& = &
 -\sum_{r=1}^N \alpha_r {1\over 1-{z_r\over z}}{z_r\over z} + o({1\over z^2})
\; = \;
-\left( \sum_{r=1}^N \alpha_r z_r\right){1\over z} + o({1\over z^2}) \ .
\end{eqnarray*}
For the radius $r_0$ of the  circle surrounding 0 on the light-cone diagram
one thus gets
$$
r_0
\; \equiv \;
|\rho(z)|
\; \approx \;
{1\over |z|}
\left|\sum_{r=1}^N \alpha_r z_r\right|
\; = \;
{A\over \epsilon_\infty}\ .
$$
In the limit $|z| \to \infty$ one also has
$$
\sigma
\; = \;
- 2 \ln |z|^2 + \ln A^2,
\hskip 5mm
n^a\partial_a \sigma
\; = \;
- {4\over \epsilon_\infty}\ ,
$$
and
$$
{1 \over 8\pi}
\int\limits_{\partial M_\infty}\!ds\, \sigma n^a\partial_a \sigma
\; = \;
4\ln\epsilon_\infty - 2\ln A
\; = \;
-4\ln r_0 + 2\ln A\ .
$$
\subsection{Identities}

Comparing the r.h.s of (\ref{global_r}), (\ref{local_r})
and (\ref{global_I}), (\ref{local_I})
one gets the identities
\begin{eqnarray}
\label{alpha_r}
- \ln |\alpha_r|
& = &
-\ln \left|\sum\limits_{r=1}^N \alpha_r z_r\right|
+\sum\limits_{s(s\neq r)} \ln |z_r-z_s|
-  \sum_{I} \ln |Z_I-z_r|\ ,\\
 \label{c_I}
 \ln c_I
& = &
\ln \left|\sum\limits_{r=1}^N\alpha_r z_r\right|
+\sum\limits_{J(J\neq I)} \ln |Z_I-Z_J|
-  \sum_{r} \ln |z_r-Z_I|\ .
\end{eqnarray}
Summing formula (\ref{alpha_r}) over $r$
and formula (\ref{c_I}) over $I$
and adding the results  one gets the equality
\begin{eqnarray*}
&&
\hskip -5mm
\sum\limits_{I=1}^{N-2}\ln c_I - \sum\limits_{r=1}^N \ln|\alpha_r|
\; =\; \\
&&
\hskip 1cm = \; -\ln \left|\sum\limits_{r=1}^N \alpha_r z_r\right|^2
+ \sum\limits_{s< r} \ln |z_r-z_s|^2
 -  \sum_{I,r} \ln |Z_I-z_r|^2
+ \sum\limits_{J< I} \ln |Z_I-Z_J|^2\ .
\end{eqnarray*}
Introducing compact notation
\begin{eqnarray}
\label{notation}
 [c] & \equiv & \prod\limits_I c_I\ , \hskip 1cm
{[\alpha]} \; \equiv \;   \prod\limits_{r=1}^N \alpha_r\ , \hskip
1cm {[\epsilon]} \; \equiv \;  \prod\limits_{r=1}^N \epsilon_r\ ,
\\
 {[P]}
& \equiv & { \prod\limits_{1\leqslant r<s\leqslant N}|z_r-z_s|
\prod\limits_{ I<J} |Z_I-Z_J|\over \prod\limits_{r,I} |z_r-Z_I|}\
, \nonumber
\end{eqnarray}
it can be written in the form
\begin{equation}
\label{identity}
[c][\alpha]^{-1}= A^{-2}[P]^2\ .
\end{equation}

\subsection{Conformal anomaly}

We calculate the Liouville action for the conformal factor of the Mandelstam map
\begin{eqnarray*}
S_L[g,\sigma]&\equiv&
{1 \over 8\pi} \int\! \sqrt{g} d^2z\,
 (g^{ab} \partial_a \sigma \partial_b \sigma + 2 R_g \sigma )
+ {1\over 2\pi} \int\! ds\, \kappa_g \sigma\\
& = &
{1 \over 8\pi} \sum_{r=1}^N
\int\limits_{\partial M_r}\!ds\, \sigma n^a\partial_a \sigma
+{1\over 2\pi}\sum_{r=1}^N \int\limits_{\partial M_r}\!ds\, \kappa_g \sigma \\
& + &
{1 \over 8\pi} \sum_{I=1}^{N-2}
\int\limits_{\partial M_I}\!ds\, \sigma n^a\partial_a \sigma
+{1 \over 8\pi}
\int\limits_{\partial M_\infty}\!ds\, \sigma n^a\partial_a \sigma\ .
\end{eqnarray*}
The last line of the formula above is necessary for regularization of the
singularities arising at the points $Z_I$ and $z \to \infty$ which correspond
via the full Mandelstam map to internal points on the light-cone diagram.

The geodesic curvature on small circles surrounding $z_r$ is
$\kappa_g = -{1\over \epsilon_r}$
and
$$
{1\over 2\pi}\int\limits_{\partial M_r} ds \kappa_g \sigma =
-2\ln |\alpha_r|+2\ln\epsilon_r\ .
$$
Using expressions (\ref{local_r}), (\ref{local_I}) and (\ref{local_0})
for other boundary integrals one gets
\begin{eqnarray*}
S_L[g,\sigma]
& = &
-\sum_{r=1}^N \ln|\alpha_r|+ \sum_{r=1}^N\ln\epsilon_r
- {1 \over 2} \sum_{I=1}^{N-2}\ln c_I
+ 2\ln A
-4\ln r_0
- {1 \over 2} \sum_{I=1}^{N-2}\ln 2r_I
\end{eqnarray*}
and
\begin{eqnarray*}
\left({\rm det} {\cal L}_{{\rm e}^\sigma g} \right)^{-{1\over 2}}
& = &
{\rm e}^{{1\over 12}S_L[g,\sigma]}
\left({\rm det} {\cal L}_{g} \right)^{-{1\over 2}} \\
& = &
[\epsilon]^{{1\over 12}}
A^{{1\over 6}}
[\alpha]^{-{1\over 12}} [c]^{-{1\over 24}}
\left({\rm det} {\cal L}_{ g} \right)^{-{1\over 2}} \ .
\nonumber
\end{eqnarray*}
Using  formula (\ref{identity}) it can be written in the form
\begin{eqnarray}
\label{ano}
\left({\rm det} {\cal L}_{{\rm e}^\sigma g} \right)^{-{1\over 2}}
& = &
[\epsilon]^{{1\over 12}}
A^{{1\over 4}}
[\alpha]^{-{1\over 8}} [P]^{-{1\over 12 }}
\left({\rm det} {\cal L}_{ g} \right)^{-{1\over 2}} \ .
\end{eqnarray}

\subsection{Jacobian for the Mandelstam map}

Interaction point $\rho_I$ depends on $z_r$ through the relation
\begin{equation}
\label{rhoint_def}
\rho_I \; = \; \sum\limits_{r}\alpha_r\ln(Z_I-z_r)\ ,
\end{equation}
where $Z_I= Z_I(\alpha_s,z_s)$ is a solution of the equation
\begin{equation}
\label{zint_def}
\sum\limits_{r}\frac{\alpha_r}{Z_I-z_r} \; = 0\; \ .
\end{equation}
When we change the integration variables from
differences of the interaction points $\rho_I-\rho_1$
to $z_s-$s (with $I, s = 2 ,\ldots,N-2$), the corresponding
Jacobian reads
$$
{\cal J} \; = \;
\left|\det\left(\frac{\partial(\rho_I-\rho_1)}{\partial z_s}\right)\right|^2 .
$$

Differentiating (\ref{rhoint_def}) with respect to $z_s$ and using (\ref{zint_def}) we get
$$
\frac{\partial\rho_I}{\partial z_s} \; = \;
 \left(\sum\limits_{r=1}^{N-1}\frac{\alpha_r}{Z_I-z_r}\right)\frac{\partial Z_I}{\partial z_s}
- \sum\limits_{r=1}^{N-1}\frac{\alpha_r}{Z_I-z_r}\ \frac{\partial z_r}{\partial z_s}
\; = \;
- \frac{\alpha_s}{Z_I-z_s}\ ,
$$
and consequently
$$
{\cal J}
\;= \;
\left|\det\left(\frac{\alpha_s(Z_I-Z_1)}{(Z_I-z_s)(Z_1 - x_s)}\right)\right|^2
\;  = \;
\prod\limits_{r=2}^{N-2}\left|\frac{\alpha_r}{Z_1-z_r}\right|^2
\prod\limits_{J=2}^{N-2}\left|Z_J-Z_1\right|^2
\ \left|\det\,C\right|^2 ,
$$
where $C$ is an $(N-3)\times(N-3)$ matrix with elements
$$
C_{Ir} \; = \; \frac{1}{Z_I-z_r}\ .
$$
The identity
\begin{equation}
\label{det1} {\rm det}\ C \; = \; w_{N-3}\!\!\!\!\!\!\! \prod_{2
\leqslant r < s \leqslant N-2}\!\!\!\!\!\!\!(z_s-z_r) \prod_{2
\leqslant I<J}\!\!(Z_J-Z_I) \prod_{r,I = 2}^{N-2}(Z_I-z_r)^{-1}\ ,
\end{equation}
where $w_{N-3}$ depends only on $N,$ follows from the fact that
both sides of (\ref{det1}) are analytic functions of all $z_r-$s and $Z_I-$s with the
same locations and orders of zeroes and poles.
If we choose
$Z_I = I$ and $ z_r = 1-r,$ then
the l.h.s. of the Eq. (\ref{det1}) reduces to the special case of the
Hankel's de\-ter\-mi\-nant
$$
{\rm det}\left(
\begin{array}{cccc}
1 & 1/2 & \ldots & 1/n \\
1/2 & 1/3 & \ldots &1/(n+1)  \\
\vdots &&& \\
1/n & 1/(n+1) & \ldots & 1/(2n-1)
\end{array}
\right)
\; = \;
\left(\prod_{k=1}^{n-1}k!\right)^3\
    \left[\prod_{k=1}^{n}(n+k-1)!\right]^{-1}
$$
with $n=N-3,$ while the r.h.s. is equal to
$$
w_n\ (-1)^{\frac12n(n-1)}\left(\prod_{k=1}^{n-1}k!\right)^3\
    \left[\prod_{k=1}^{n}(n+k-1)!\right]^{-1}\ .
$$
This leads to
$$
w_{N-3} \; = \; (-1)^{\frac12(N-3)(N-4)}
$$
and gives
\begin{eqnarray}
\label{jacob_1} {\cal J} & = &
\prod\limits_{r=2}^{N-2}|\alpha_r|^2 \!\!\!\!\! \prod\limits_{2
\leqslant r < \leqslant N-2}\!\!\!\!\!\!|z_r-z_s|^2  \!\!\!\!
\prod\limits_{1 \leqslant I < J}\!|Z_I-Z_J|^2
\prod\limits_{r=2}^{N-2}\prod\limits_{I=1}^{N-2}|Z_I-z_r|^{-2}
\nonumber  \\
& = & {[\alpha]}^2{[P]}^2\left|(z_N-z_{N-1})(z_N - z_1)(z_{N-1}-z_1)\right|^2
\!\!\!\!
\prod\limits_{t=1,N-1,N}\left\{\frac{1}{|\alpha_t|^2}\frac{\prod_I|Z_I-z_t|^2}
{\prod_{s\neq t}|z_s-z_t|^2}\right\}
\nonumber \\
& = &
\left|(z_N-z_{N-1})(z_N - z_1)(z_{N-1}-z_1)\right|^2  \
A^{-6}{[\alpha]}^2{[P]}^2\ ,
\end{eqnarray}
where the last equality follows from (\ref{alpha_r}) and $A$ is defined by (\ref{A}).

\renewcommand{\thesection}{B.}
\renewcommand{\thesubsection}{B.\arabic{subsection}}
\renewcommand{\theequation}{B.\arabic{equation}}
\setcounter{equation}{0}

\section{Operator--states correspondence}
Let $C_r$ stands for a part o the world-sheet of $r^{\rm th}$
external string,
$$
C_r\; = \; \left\{\rho \; : \;\frac{\tau_r}{\alpha_r}\ \geqslant\
\Re\,\frac{\rho}{\alpha_r}\ \geqslant\
\frac{\tau_r'}{\alpha_r}\right\}\ .
$$
Using the equivalence of the operator and path integral
expression for the evolution operator between light-cone times
$\tau_r'$ and $\tau_r$ one can write
\begin{eqnarray}
\label{amplit_3}
  \Psi_r[X^i_r(\sigma_r)] & = & \int_{\tau_r'}\!{\cal D}X^i_{r'}(\sigma_r')\;
  \int_{C_r}\!\!{\cal D}^{g_{\scriptscriptstyle\rm lc}}X^i(\sigma,\tau)\;
  {\rm e}^{-S[g_{\scriptscriptstyle\rm lc},X^i(\sigma,\tau)]}\;
  {\rm e}^{-\frac{1}{2\sqrt\alpha}(P^{\rm\scriptscriptstyle T})_r^{-}
  (\tau_r'-\tau_r)}\Psi_r[X^i_{r'}(\sigma_r)] \nonumber \\
  && \\
  & = & {\rm e}^{\frac{\lambda_r^{\rm\scriptscriptstyle T}}{\alpha_r}
  (\tau_r-\tau_r')}\int_{\tau_r'}\!{\cal D}X^i_{r'}(\sigma_r')\;
  \int_{C_r}\!\!{\cal D}^{g_{\scriptscriptstyle\rm lc}}X^i(\sigma,\tau)\;
  {\rm e}^{-S[g_{\scriptscriptstyle\rm lc},X^i(\sigma,\tau)]}\;
  \Psi_r[X^i_{r'}(\sigma_r)]\ , \nonumber
\end{eqnarray}
where the inner path integral goes over the fields satisfying the
Dirichlet boundary conditions
$$
X^i(\sigma_r,\tau_r) \; = \; X^i_r(\sigma_r)\ , \hskip 1cm
X^i(\sigma_r,\tau_r') \; = \; X^i_{r'}(\sigma_r)\ .
$$
The functional Fourier transform at the light-cone time $\tau_r'$
allows to convert the wave functional $\Psi_r[X^i(\sigma_r)]$ to
the momentum representation and Eq. (\ref{amplit_3}) takes the
form
\begin{eqnarray}
\label{amplit_4}
 \Psi_r[X^i_r(\sigma_r)] & = &
 {\rm e}^{\frac{\lambda_r^{\rm\scriptscriptstyle T}}{\alpha_r}
 (\tau_r-\tau_r')}
 \int_{C_r}\!\!{\cal D}^{g_{\scriptscriptstyle\rm lc}}X^i(\sigma,\tau)\;
  {\rm e}^{-S[g_{\scriptscriptstyle\rm lc},X^i(\sigma,\tau)]}\
  \times
  \\ &&
  \hskip 1cm \times \ \int_{\tau_r'}\!{\cal D}P^i_{r'}(\sigma_r')\;
  {\rm e}^{i\int\!d\sigma_r'\,
  P^i_r(\sigma_r')X^i(\sigma_r',\tau_r')}\;
  \Psi_r[P^i_{r'}(\sigma_r)]\ , \nonumber
\end{eqnarray}
with
$$
X^i(\sigma_r,\tau_r) \; = \; X^i_r(\sigma_r)\ , \hskip 1cm
\left.\partial_\tau X^i(\sigma_r,\tau)\right|_{\tau = \tau_r'} \;
= \; 0\ .
$$
Using the conformal map
\begin{equation}
\label{rho_0} \rho_r(z) = \alpha_r\ln(z-z_r)
\end{equation}
one can rewrite Eq. (\ref{amplit_4}) in terms of the path integral
on  the annulus
$$
P_r =\left\{\eta_r'= {\rm e}^{{\tau'_r \over \alpha_r}}\leqslant |z-z_r|\leqslant
\eta_r= {\rm e}^{{\tau_r \over \alpha_r}}\right\}
$$
 in the complex plane endowed with the
standard flat metric $g_{\rm pl}$:
\begin{eqnarray}
\label{amplit_5}
 \Psi_r[X^i_r(\sigma_r)] & = &
 {\rm e}^{
{\lambda_r^{\rm\scriptscriptstyle T}\over \alpha_r}
(\tau_r-\tau_r') }
 \left({\cal A}_{P_r}\right)^{c^{\rm\scriptscriptstyle T}}
 \int_{P_r}\!\!{\cal D}^{g_{\scriptscriptstyle\rm pl}}
  X^i(w,\bar w)\;
  {\rm e}^{-S[g_{\scriptscriptstyle\rm pl},X^i(w,\bar w)]}\;
   \; \times \\
  &  \times&
  \int_{\tau_r'}\!{\cal D}P^i_{r'}(\sigma_r')\;
  {\rm e}^{i\int\!d\sigma_r\;P^i_{r'}(\sigma_r')
  X^i(z_r +{\rm e}^{\frac{\tau_r'+ i\sigma_r'}{\alpha_r}},
  \bar z_r +{\rm e}^{\frac{\tau_r'- i\sigma_r'}{\alpha_r}})}\
  \Psi_r[P^i_{r'}(\sigma_r')]\ ,
  \nonumber
\end{eqnarray}
where the conformal anomaly factor reads
$$
{\cal A}_{P_r} = {\rm
e}^{{\tau_r'-\tau_r\over 12 \alpha_r}} \ .
$$
In the limit $\eta'_r\to 0$ one can replace the integration over
the fields defined on the annulus $P_r$ with the boundary
conditions on the ``small'' boundary determined by the state
functional $\Psi_r[P^i_{r'}(\sigma_r')]$, by the integration over
the fields on the disc $D_r=\{z:|z-z_r|\leqslant \eta_r\}$ with an
appropriate operator insertion $V_r^{\rm\scriptscriptstyle
T}(z_r,\bar z_r)$ at $z_r$. In this limit the $\eta'_r$-dependence
drops out. Indeed the contribution from the partition function on
the annulus:
$$
 \int_{P_r}\!\!{\cal D}^{g_{\scriptscriptstyle\rm pl}}
  X^i(w,\bar w)\;
  {\rm e}^{-S[g_{\scriptscriptstyle\rm pl},X^i(w,\bar w)]}
\propto (\eta_r')^{-{c^{\rm\scriptscriptstyle T}\over 6}} \ ,
  $$
and  the ``self interaction'' term $(\eta_r')^{\Delta_r}$ resulting
from the singular behavior of the classical action are cancelled
by the term $(\eta_r')^{-(\lambda_r^{\rm\scriptscriptstyle T}-
\frac{c^{\rm\scriptscriptstyle T}}{12})}$. One thus gets
\begin{equation}
\label{amplit_6}
\Psi_r[X^i_r(\sigma_r)]  \; = \; {\rm
e}^{(\lambda_r^{\rm\scriptscriptstyle T}-
\frac{c^{\rm\scriptscriptstyle T}}{12})
\frac{\tau_r}{\alpha_r}}
 \int_{D_r}\!\!{\cal D}^{g_{\scriptscriptstyle\rm pl}}
 X^i(w,\bar w)\;
  {\rm e}^{-S[g_{\scriptscriptstyle\rm pl}, X^i(w,\bar w)]}
  \ V_r^{\rm\scriptscriptstyle T}(z_r,\bar z_r)\ .
\end{equation}

\renewcommand{\thesection}{C.}
\renewcommand{\thesubsection}{C.\arabic{subsection}}
\renewcommand{\theequation}{C.\arabic{equation}}
\setcounter{equation}{0}

\section{Calculation of the contour integrals}

To calculate the integral
\begin{eqnarray*}
 \Upsilon_I^i{}^{(1)}
 & \equiv &
 \oint\limits_{\rho^{-1}(C_I)}\frac{dz}{\pi i}\;\frac{1}{\rho'(z)}
{\cal T}\left[T^{\rm\scriptscriptstyle T}(z)\left(\chi^i(z)-\chi^i(z_I)\right)\right] \\
& = & \lim_{\varepsilon\to 0}\oint\limits_{z_I}\frac{dz}{\pi i}\;
\left(\frac{d\rho(w)}{dw}\right)^{-1} {\cal
T}\left[T^{\rm\scriptscriptstyle
T}(w)\left(\chi^i(z)-\chi^i(z_I)\right)\right]\ ,
\end{eqnarray*}
where
\begin{equation}
\label{z1def} \varepsilon \; = \; \rho(w) - \rho(z)
\end{equation}
is small compared to $z-z_I,$ we use the identity
$$
T^{\rm\scriptscriptstyle T}(w)\chi^i(z) \; = \;
\frac{1}{w-z}\partial_w\chi^i(w) + :T^{\rm\scriptscriptstyle
T}(w)\chi^i(z)\!: \;\; = \; \frac{1}{w-z}\partial_w\chi^i(w) +
:T^{\rm\scriptscriptstyle T}(z)\chi^i(z)\!: + {\cal
O}(\varepsilon)\ .
$$
It gives
\begin{eqnarray*}
\Upsilon_I^i{}^{(1)}
 & = & \lim_{\varepsilon\to
0}\oint\limits_{z_I}\frac{dz}{\pi i}\;
\left(\frac{d\rho(w)}{dw}\right)^{-1}
\left(\frac{1}{w-z}-\frac{1}{w-z_I}\right)\partial_w\chi^i(w)  \\
& + & \oint\limits_{z_I}\frac{dz}{\pi i}\;
\left(\frac{d\rho(z)}{dz}\right)^{-1} :T^{\rm\scriptscriptstyle
T}(w)\left(\chi^i(z)-\chi^i(z_I)\right)\!:\ .
\end{eqnarray*}
The second integral vanishes due to the zero of
$\chi^i(z)-\chi^i(z_I)$ for $z\to z_I.$

Differentiation of (\ref{z1def}) with respect to $z$ gives
$$
\frac{d\rho(w)}{dz} \; = \; \frac{d\rho(z)}{dz}\equiv \rho'(z)\ ,
$$
and
$$
\frac{d\rho(w)}{dw} \; = \;
\frac{d\rho(w)}{dz}\left(\frac{dw}{dz}\right)^{-1} \; = \;
\rho'(z)\left(\frac{dw}{dz}\right)^{-1}.
$$
Expanding  $\rho(w)$  around $z$ and inverting the resulting
series one gets
$$
w - z \; = \; \frac{1}{\rho'(z)}\,\varepsilon - \frac{\rho''(z)}
{2\left(\rho'(z)\right)^2}\,\varepsilon^2 + {\cal
O}\left(\varepsilon^3\right)\ .
$$
Consequently
\begin{eqnarray*}
\frac{1}{w-z} & = & \frac{\rho'(z)}{\varepsilon} +
\frac12\,\frac{\rho''(z)}{\rho'(z)} + {\cal
O}\left(\varepsilon\right)\ ,
\\
\frac{dw}{dz} & = & 1 -
\frac{\rho''(z)}{\left(\rho'(z)\right)^2}\,\varepsilon +
{\cal O}\left(\varepsilon^2\right)\ , \\
\left(\frac{d\rho(w)}{dw}\right)^{-1}\frac{1}{w-z} & = &
\frac{1}{\varepsilon} -
\frac12\frac{\rho''(z)}{\left(\rho'(z)\right)^2} + {\cal
O}\left(\varepsilon\right)\ ,
\end{eqnarray*}
and
\begin{equation}
\label{Gamma1_2} \Upsilon_I^i{}^{(1)}
 \; = \;
  - \oint\limits_{z_I}\frac{dz}{\pi i}\;\frac{1}{\rho'(z)}
  \left(\frac12\frac{\rho''(z)}{\rho'(z)} + \frac{1}{z-z_I}
 \right)\partial_z\chi^i(z)\ ,
\end{equation}
where we have used the fact that $\partial_z\chi^i(z)$ is regular
at $z=z_I$ so that
$$
\frac{1}{\varepsilon} \oint\limits_{z_I}\frac{dz}{\pi
i}\;\partial_z\chi^i(z) \; = \; 0
$$
for any $\varepsilon.$

If we write Taylor expansion of $\rho(z)$ around $z_I$ as
$$
\rho(z) - \rho(z_I) = \frac12a(z-z_I)^2 + \frac13b(z-z_I)^3 +
{\cal O}\left((z-z_I)^4\right)\ ,
$$
then from (\ref{Gamma1_2})
$$
\Upsilon_I^i{}^{(1)}
 \; = \;\frac{2b}{a^2}\ \partial\chi^i(z_I) -
\frac{3}{a}\ \partial^2\chi^i(z_I) \ ,
$$
where $\partial\chi^i(z_I) \equiv \partial_{z_I}\chi^i(z_I).$ It
is also immediately to check that
$$
  \oint\limits_{z_I}\frac{dz}{\pi i}\;\frac{1}{\rho'(z)} \left(
\frac{\rho^{(3)}(z)}{\rho'(z)}-\frac32\frac{\left(\rho''(z)\right)^2}{\left(\rho'(z)\right)^2}
\right)\left(\chi^i(z) - \chi^i(z_I)\right) \\
 \; = \;  \frac{b}{a^2}\
\partial\chi^i(z_I)- \frac{3}{2a}\ \partial^2\chi^i(z_I) \ ,
$$
what finally gives
$$
 \Upsilon_I^i
 \; = \;
 \left(1- \frac{c^{\rm \scriptscriptstyle T} +c^{\rm \scriptscriptstyle B}
 +c^{\rm \scriptscriptstyle L}}{24}\right) \frac{2b}{a^2}\
 \partial\chi^i(z_I) + \left(\frac{c^{\rm \scriptscriptstyle
 T}+c^{\rm \scriptscriptstyle B} +c^{\rm \scriptscriptstyle
 L}}{16}-\frac32\right) \frac{2}{a}\
 \partial^2\chi^i(z_I)\ .
$$

\vskip 3mm
\noindent
{\Large\bf Acknowledgements}

\vskip 2mm

L.H. would like to thank the faculty of the C.~N.~Yang Institute
of Theoretical Physics for their warm hospitality during his stay
at Stony Brook, where part of this work has been completed. The
work of L.H. was partially supported by the Foundation for the
Polish Science fellowship.

Z.J. is grateful to R.~Flume and R.~Poghossian for illuminating
discussions. He   thanks the Physics Institute of Bonn University
for its hospitality during his stay at Bonn, where part of this
work was done. The financial support of DAAD foundation for this
stay is acknowledged.

\end{document}